\documentclass[twocolumn,prb,showpacs,superscriptaddress,preprintnumbers,amsmath,amssymb]{revtex4}
\usepackage{graphicx}
\usepackage{mathptmx}
\usepackage{verbatim}
\usepackage{epstopdf}
\usepackage{amssymb}
\usepackage{mathdots}
\usepackage{amsfonts}
\usepackage{mathrsfs}
\usepackage{amsmath}  
\usepackage{color}
\usepackage{rotating}
\usepackage[T1]{fontenc}
\usepackage{bm}  
\usepackage{xspace}
\usepackage{dcolumn}  

\newcommand{\angstrom}{\textup{\AA}} 

\begin{document}
\title{
Carrier and strain tunable intrinsic magnetism in two-dimensional MAX$_3$ transition metal chalcogenides
}
\author{Bheema Lingam Chittari}
\affiliation{Department of Physics, University of Seoul, Seoul 02504, Korea}
\affiliation{SKKU Advanced Institute of Nanotechnology, Sungkyunkwan University, Suwon, 16419, Korea}
\author{Dongkyu Lee\footnote{Present address: Department of Physics and Photon Science, School of Physics and Chemistry, Gwangju Institute of Science and Technology, Gwangju 61005, Korea}}
\affiliation{Department of Physics, University of Seoul, Seoul 02504, Korea}
\author{Allan H. MacDonald}
\affiliation{Department of Physics, The University of Texas at Austin, Austin, Texas 78712, USA }
\author{Euyheon Hwang}
\email{euyheon@skku.edu}
\affiliation{SKKU Advanced Institute of Nanotechnology, Sungkyunkwan University, Suwon, 16419, Korea}
\author{Jeil Jung}
\email{jeiljung@uos.ac.kr}
\affiliation{Department of Physics, University of Seoul, Seoul 02504, Korea}

\begin{abstract}
We present a density functional theory study of the carrier-density and strain dependence   
of magnetic order in two-dimensional (2D) MAX$_3$ (M= V, Cr, Mn, Fe, Co, Ni; A= Si, Ge, Sn, and X= S, Se, Te)
transition metal trichalcogenides.  
Our {\em ab initio} calculations show that this class of compounds 
includes wide and narrow gap semiconductors and metals and half-metals, 
and that most of these compounds are magnetic.
Although antiferromagnetic order is most common,
ferromagnetism is predicted in 
MSiSe$_3$ for M= Mn, Ni, in MSiTe$_3$ for M= V, Ni,
in MnGeSe$_3$, in MGeTe$_3$ for M=Cr, Mn, Ni, in FeSnS$_3$,
and in MSnTe$_3$ for M= V, Mn, Fe.  Among these compounds 
CrGeTe$_3$ and VSnTe$_3$ are ferromagnetic semiconductors.
Our calculations suggest that the competition between antiferromagnetic and ferromagnetic order can be substantially altered by 
strain engineering, and in the semiconductor case also by gating. 
The associated critical temperatures can be substantially enhanced by means of carrier doping and strains. 
\end{abstract}


\maketitle
\section{Introduction} 
Recently interest in 2D materials research has expanded beyond 
graphene\cite{novoselov_nature,philipkim_nature} to include other layered van der Waals materials.~\cite{novoselov_pnas}
For example ultrathin transition metal dichalcogenides (TMDC),\cite{chowalla} a family
that includes metals, semiconductors with exceptionally strong 
light-matter coupling,~\cite{lightmatter} charge density waves, and superconductors,~\cite{tas2,tmdneto,frindt,mos2sc}
have emerged as a major research focus.  %
Two-dimensional (2D) materials with room-temperature magnetic order are of particular interest 
because they appear to be potentially attractive
hosts for non-volatile information storage and logic devices.
Unfortunately single-layer magnetism has so far been
realized only in 
relatively fragile 2D materials,~\cite{mpx3other16c,xiaodong2017,zhangxiang2017,hanwei2017} 
and no 2D materials have yet been discovered that exhibit room-temperature magnetism.

Recent density functional theory (DFT) studies
have proposed several potentially magnetic single-layer van der Waals materials, 
including the group-V based dichalcogenides VX$_2$ (X= S, Se),~\cite{2dmagnet1, 2dmagtmd}
FeBr$_3$, chromium based ternary tritellurides CrSiTe$_3$ and CrGeTe$_3$,~\cite{lebegue, max1, max2, max3, max4, max5, max6,sivadas,CGeT-PRB2017, zhangxiang2017, CGeT-ChemMat2017, CGeT-ChemMat2016, CGeT-JJAP2016, CGeT-PRB2017-2, CGeT-2DMat2017, MnST-PRB2017, CrST-SciRep2016, CrSGT-PRL2016, MnPSe-PRL2016, CGeT-2DMat2016}
and MnPX$_3$  ternary chalcogenides.~\cite{niu,jacs}
Separately the  CrATe$_3$ (A= Si, Ge)~\cite{max1} ternary tritellurides have been predicted
by local density approximation (LDA) density functional theory (DFT) calculations
to be ferromagnetic semiconductors with small band gaps of 0.04 and 0.06 eV respectively.
The few layer limits of these materials have been studied 
experimentally by performing temperature dependent transport~\cite{max6},
and neutron scattering experiments, and these have been interpreted as 
providing suggestions of 2D magnetism.~\cite{max3}  
In its single layer limit CrSiTe$_3$ is found to be a semiconductor with a
generalized gradient approximation (GGA) 
gap of 0.4~eV,~\cite{max5} substantially larger than its bulk value, and to have
negative thermal expansion.~\cite{max2} 
Different authors have reached various conclusions concerning the 
relative stability between ferromagnetic~\cite{max1,max2,max5} and 
antiferromagnetic phases,~\cite{sivadas}
reflecting a relatively small total energy differences between the energies of the
two magnetic phases. For compounds like CrGeTe$_3$ and CrSnTe$_3$ involving larger atomic 
number Ge and Sn atoms,
DFT predicts ferromagnetic semiconducting phases
with Curie temperatures between 80-170K.~\cite{sivadas,max4,max6}
Among non-chalcogenide compounds the CrX$_3$ (X= F, Cl, Br, I)~\cite{cri3,crx3} trihalides 
are expected to be ferromagnetic semiconductors, with Curie temperatures T$_{\rm C} < 100$~K.
A Recent breakthrough experiment has realized CrI$_3$ based devices in ultrathin layered form 
and demonstrated an intricate competition between ferromagnetic (FM) and antiferromagnetic (AFM) 
states as a function of layer number and external fields~\cite{xiaodong2017}. 

In this paper we present an exhaustive DFT survey of the magnetic phases  
of single-layer transition metal trichalcogenide compounds of the MAX$_3$ family.
Our survey covers a variety of late 3d transition metals (M= V, Cr, Mn, Fe, Co, Ni),
the group IV elements (A=  Si, Ge, Sn), and the three chalcogen atoms (X= S, Se, Te). 
These single-layer compounds are structurally closely related to their 
transition metal trichalcogenide MPX$_3$  compound cousins,
which we studied in a recent work~\cite{Bheema_PRB}.
The main difference between MAX$_3$ compounds and 
MPX$_3$ compounds is that the group V phosphorus (P) atom inside the
(P$_2$X$_6)^{4-}$ skeleton is replaced by  (A$_2$X$_6$)$^{6-}$ bipyramid ions with X = (S, Se, Te) group IV elements.
We find that this change is responsible for important modifications in the resulting electronic and magnetic properties.
We have examined how the electronic structures are modified when the magnetic phases
changes from antiferromagnetic to ferromagnetic, or from magnetic to non-magnetic, 
when we modify the electron carrier density, and when strains are applied.
Our results indicate a strong interdependence between magnetism and structural 
properties in MAX$_3$ compounds due to a surprisingly strong dependence of exchange interaction 
strengths on electron densities and strains.
Because these materials involve transition metals and therefore have strong correlations,
we do not expect quantitatively reliable predictions of density functional theory for all properties.  
Our goal with this survey is to provide insight into expected materials-property trends, and to examine the possibility of
engineering magnetic properties in these materials using field effects and strains.

Our paper is structured as follows.
We begin in Sec.~\ref{sec:methods} with a summary of some technical details of our first principles electronic structure calculations.
In Sec.~\ref{sec:groundstate} we discuss our results for ground-state properties including structure, magnetic properties, 
and electronic band structures and densities-of-states. 
Sec.~\ref{sec:tunability} is devoted to an analysis of the carrier-density dependence of the magnetic ground states,
and to the influence of strains on the magnetic phase diagram. 
Finally in Sec.~\ref{sec:summary} we present a summary and discussion of our results. 
\begin{figure}[htb!]
\begin{center}
\includegraphics[width=\columnwidth]{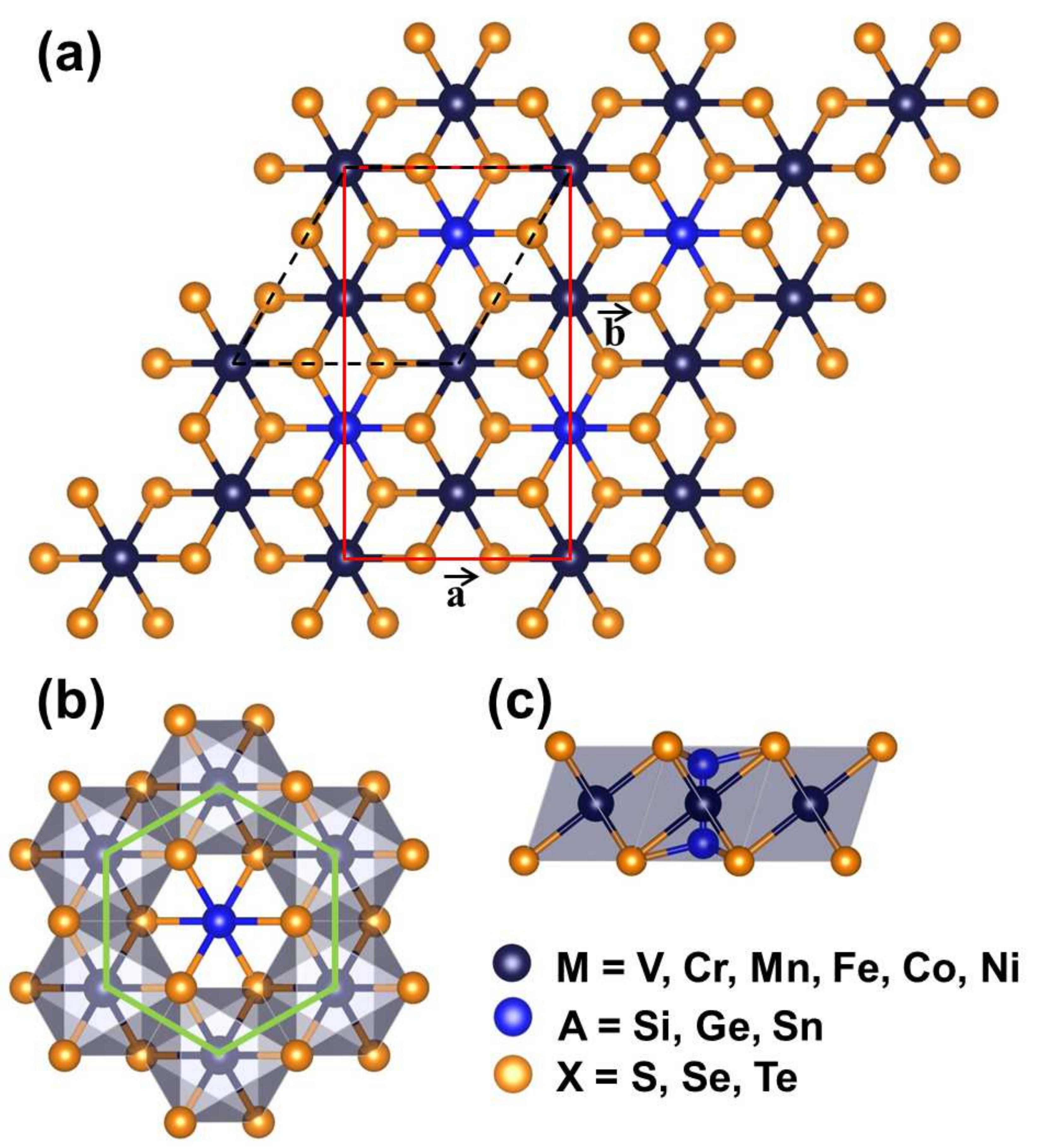}
\caption{(Color online) 
Schematic representation of the atomic structure of MAX$_3$ compounds.
{\bf a.} Atomic structure of a MAX$_3$ monolayer representing two possible choices for the unit cells,
a rectangular cell represented with red lines containing four transition metal atoms, 
and a smaller triangular cell represented by black dashed lines containing 
two transition metal atoms (M) arranged in a honeycomb. 
{\bf b.} The M atoms form hexagons (light green lines) surrounding the A (=S, Se, Te) atoms located at the center. 
{\bf c.} A side view of MAX$_3$ compounds.  
The MAX$_3$ compounds have one less occupied band than MPX$_3$ compounds because instead of the P atom pairs per unit cell 
we have group IV atoms (Si, Ge, Sn) with one fewer valence electron.  
}
\label{fig:cellstructure}     
\end{center}
\end{figure}

\section{{\em Ab initio} calculation details}
\label{sec:methods}
Ground-state electronic structure and magnetic property calculations 
have been carried out using plane-wave density functional theory as implemented in Quantum Espresso.~\cite{espresso}
We have used the Rappe-Rabe-Kaxiras-Joannopoulos ultrasoft (RRKJUS) pseudoptentials
for the semi-local Perdew-Burke-Ernzerhof (PBE) generalized gradient approximation (GGA)~\cite{GGA} together with the VdW-D2.~\cite{d2grimme} GGA+D2 (hereafter DFT-D2) wass chosen as an
electronic properties reference because of the  
overall improvement of the GGA over the LDA~\cite{LDA} for the description of in-plane covalent bonds,
and the interlayer binding captured by the longer ranged D2 correction.~\cite{noamarom}
We have also carried out calculations using GGA+D2+U (hereafter referred to as DFT-D2+U), 
using U$=$4~eV.  Some larger values of U were used for 
a few specific cases involving Co and Ni metals to obtain magnetic ground states. 
The structures were optimized until the forces on each atom reached 10$^{-5}$~Ry/a.u.
The self-consistency convergence for total energies was set to 10$^{-10}$~Ry. 
The momentum space integrals for rectangular unit cells were performed using a regularly spaced
4$\times$8$\times$1 k-point sampling density, and the plane wave energy cutoff 
was set to 60~Ry.  The out-of-plane vertical size of the periodic supercell was chosen to be 25~$\angstrom$,
which left a vacuum spacing of more than 10~$\angstrom$ between the two-dimensional layers. 
\begin{figure}[htb!]
\begin{center}
\includegraphics[width=\columnwidth]{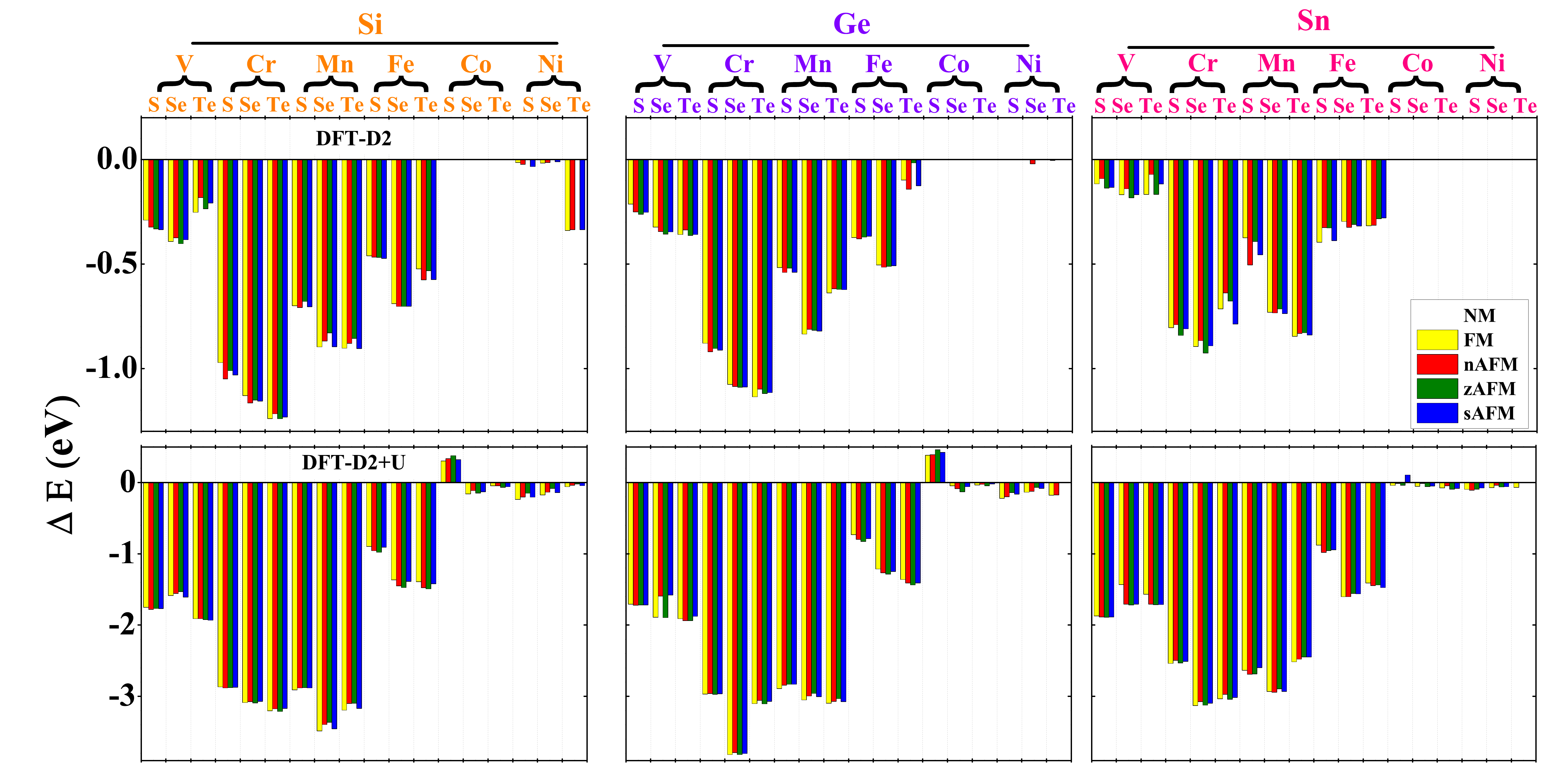}
\caption{(Color online) 
Magnetic condensation energy trends. The energy gained per metal atom due to 
magnetic order obtained within DFT-D2 and DFT-D2+U (=~4~eV). 
In the cases of CoATe$_3$, CoASe$_3$  (A = Ge, Sn), and CoSnS$_3$ we choose the U parameter 
values of 5, 6, and 7 eV respectively in order to obtain a magnetic ground-state.
Energy differences are not shown for cases in which metastable magnetic solutions could not be obtained. 
} \label{fig:groundstate}
\end{center}
\end{figure}
\begin{figure}[htb!]
\begin{center}
\includegraphics[width=\columnwidth]{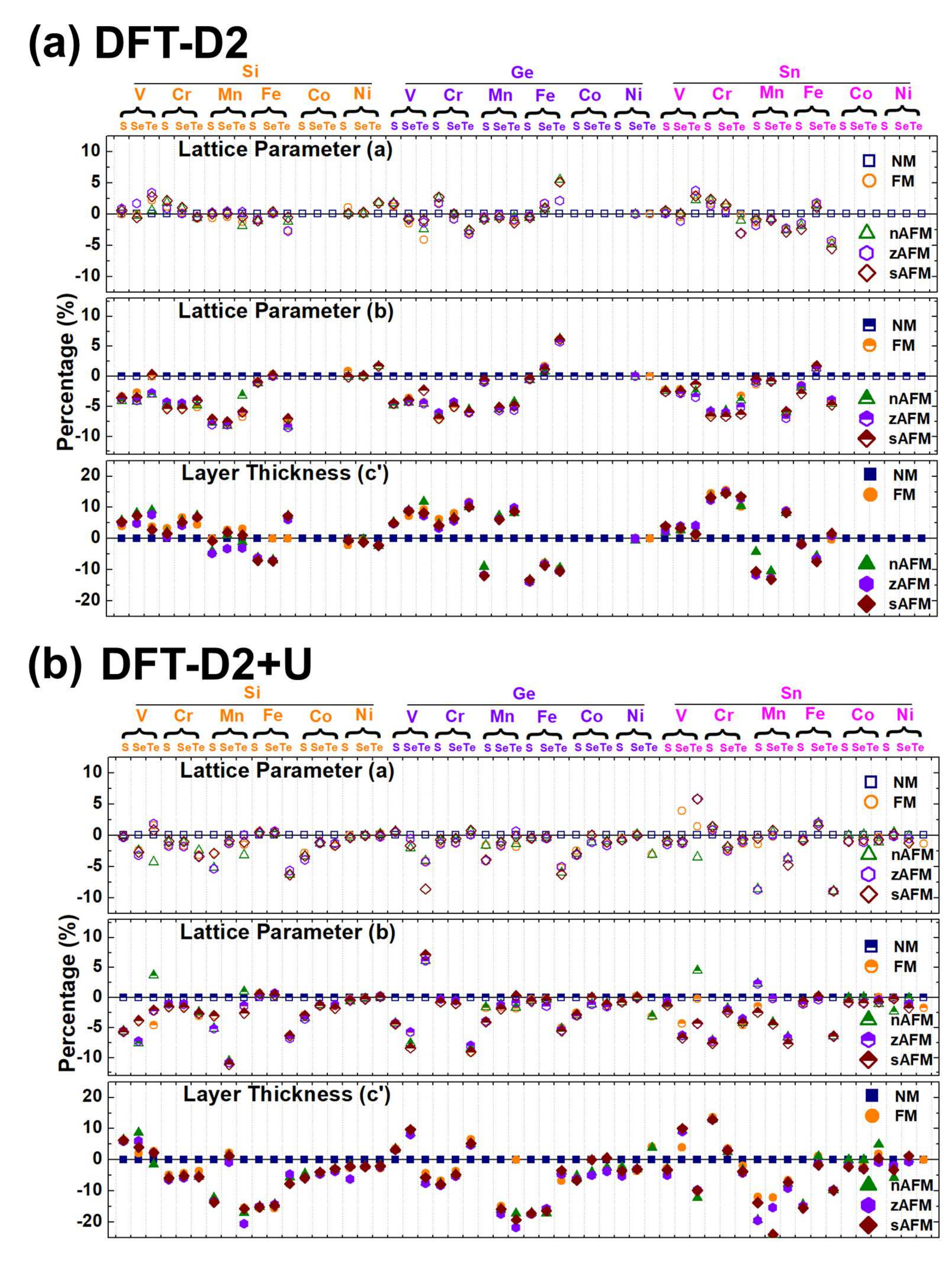}
\caption{(Color online) 
Relative distortion of the lattice parameter for different magnetic phases FM, nAFM, zAFM and sAFM
measured with respect to the structural parameters obtained for non-magnetic phases
using a rectangular unit cell with in-plane lattice parameters $a( \angstrom )$ and $b( \angstrom )$,
and layer thickness $c'( \angstrom )$.
The negative and positive values represent respectively compression and expansion during this
non-magnetic to magnetic transition.
The upper panel data a) represent calculations within DFT-D2 and the lower panel b) within DFT-D2+U. 
The latter shows generally greater variations in the lattice constants.
}
\label{fig:lattice}     
\end{center}
\end{figure}
\section{Structural and magnetic properties}
\label{sec:groundstate}
The MAX$_3$ transition metal trichalcogenide layers
consist of 3d transition metals M= (V, Cr, Mn, Fe, Co, Ni)
anchored by (A$_2$X$_6$)$^{6-}$ bipyramid ions with X = (S, Se, Te) 
and A = (Si, Ge, Sn). 
The 12 electrons taken by the six chalcogen atoms per unit cell are partly compensated by the 6 electrons provided by the sp$^3$ bonds with the bridge A atoms, leaving a final $6-$ valence state for the anionic enclosure. 
The triangular lattice of bipyramids provides enclosures 
for the transition metal atoms, forming a structure that is practically identical
to that of MPX$_3$ compounds enclosed by  (P$_2$X$_6$)$^{4-}$ bipyramids.
(See Fig.~\ref{fig:cellstructure} for a schematic illustration of the single layer unit cell ) 
The main difference is that the A atoms have one fewer electron compared to P atoms,
yielding compounds with 
larger nominal metal cation valences $3+$ rather than $2+$ in the MPX$_3$ compounds.
The interaction between the metal cations with both the chalcogen X and the bridge A atoms will determine the magnetic moment that usually concentrates at the metal atom sites.  
In our calculations all the 2D MAX$_3$ crystals we considered are magnetically ordered within DFT-D2
except for CoAX$_3$, NiGeS$_3$ and NiSnX$_3$.  These are magnetic only within DFT-D2+U 
(See Fig.~\ref{fig:groundstate} for an illustration of the trends in magnetic condensation energy,
{\it i.e.} in the energy gained by forming magnetically ordered states.). 
In the following we present an analysis of the structural, magnetic and electronic properties 
of representative 3d transition metal MAX$_3$ trichalcogenides, emphasizing 
their dependence on the chalcogen element (Si, Ge, Sn) 
and their underlying electronic band structures. 

\subsection{Structural properties}
\label{sec:structure}
In their bulk form the MAX$_3$ single layers are ABC-stacked and are held together mainly
by van der Waals forces~\cite{max1}.
Although the atomic structures of the single layer transition metal trichalcogenide crystals are similar to truncated 
bulk structures, small changes do appear due to the absence of the interlayer coupling 
and distortions in the ground-state crystal 
geometries are correlated with the magnetic phase\cite{Bheema_PRB}.  The  analysis of magnetic 
properties is simplified by the fact that the magnetic moments develop almost entirely at the metal atom sites. 
We have optimized the MAX$_3$ single layer lattice structures using the rectangular unit cell shown in Fig.~\ref{fig:cellstructure} which is 
characterized by the values of the in-plane lattice constants $a$ and $b$. 
We define the layer thickness $c'$ as the distance between the chalcogen atoms between 
the top and bottom layer in a MAX$_3$ monolayer.
The relaxed in-plane lattice parameters and layer thickness of the rectangular unit cells,
as obtained using DFT-D2 (See Tables in Supplemental Material\cite{SI}),
are found to increase for larger chalcogen atoms for given A (Si, Ge and Sn) atoms.
In general the calculated self-consistent lattice constants depend on the magnetic ordering,
The variation is substantial for the transition metals V, Cr, Mn, Fe for all combinations of A (Si, Ge, Sn)
and chalcogens S, Se, Te, up to 10\% for in-plane lattice constants and up to 20\% for the layer thickness.
For compounds with Co and Ni the lattice distortions are much smaller, see Fig.~\ref{fig:lattice}. 
As a rule of thumb we can see that the magnitude in the distortion of the bonds 
is roughly proportional to the total energy differences represented in Fig.~\ref{fig:groundstate}
and therefore they are largest when we compare magnetic and non-magnetic phases.
We have also optimized all the structures in the presence of local electron repulsion introduced 
through Hubbard U i.e. DFT-D2+U.  
This contribution leads to total energy differences between magnetic and non-magnetic phases that are roughly 
doubled (see Fig.~\ref{fig:groundstate}) within DFT-D2+U when compared to DFT-D2,
and this difference is reflected in the increase of the lattice distortions.
The relative difference of the lattice parameters between DFT-D2-U and DFT-D2 geometries are comparable
to the difference between magnetic and non-magnetic phases within the same DFT approximation, see Fig.~\ref{fig:lattice} and Fig.~1 in the Supplemental Materials.
From this observation we can expect that the short-range correlations of the transition metal atoms 
can substantially modify the energetics of the ground-state magnetic configurations.

\subsection{Magnetic properties}

\begin{figure}[htb!]
\begin{center}
\includegraphics[width=\columnwidth]{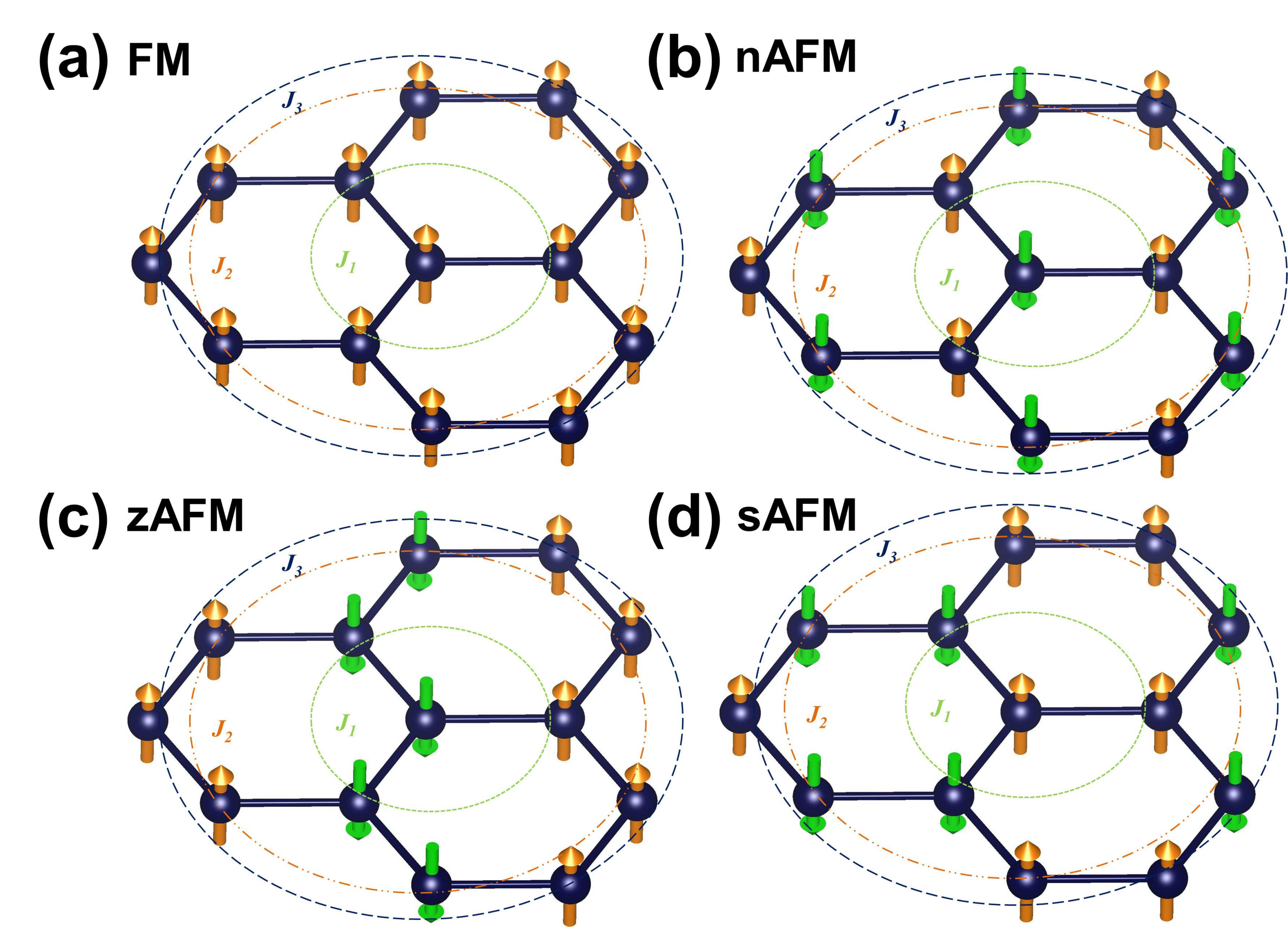}
\caption{(Color online) Schematic representations of different magnetic ordering arrangements:
(a) ferromagnetic, (b) N\'eel antiferromagnetic, (c) zigzag antiferromagnetic and (d) stripy antiferromagnetic.
In MAX$_3$ compounds, the magnetic moments reside primarily on the metal atoms which 
have a honeycomb structure in each layer. 
The red circles identify nearest neighbors (NN) of the central metal site, 
the navy blue circles (dashed-dotted) identify the second nearest neighbors, and the light blue 
circles identify the third nearest neighbors.  
The magnetic energy landscape can be approximated by 
assigning Heisenberg coupling constants ${\it J_1}$, ${\it J_2}$ and ${\it J_3}$ to metal atom pairs with
these three separations.} \label{fig:magneticorder}
\end{center}
\end{figure}

\begin{figure*}[htbp!]
\begin{center}
\includegraphics[width=18cm]{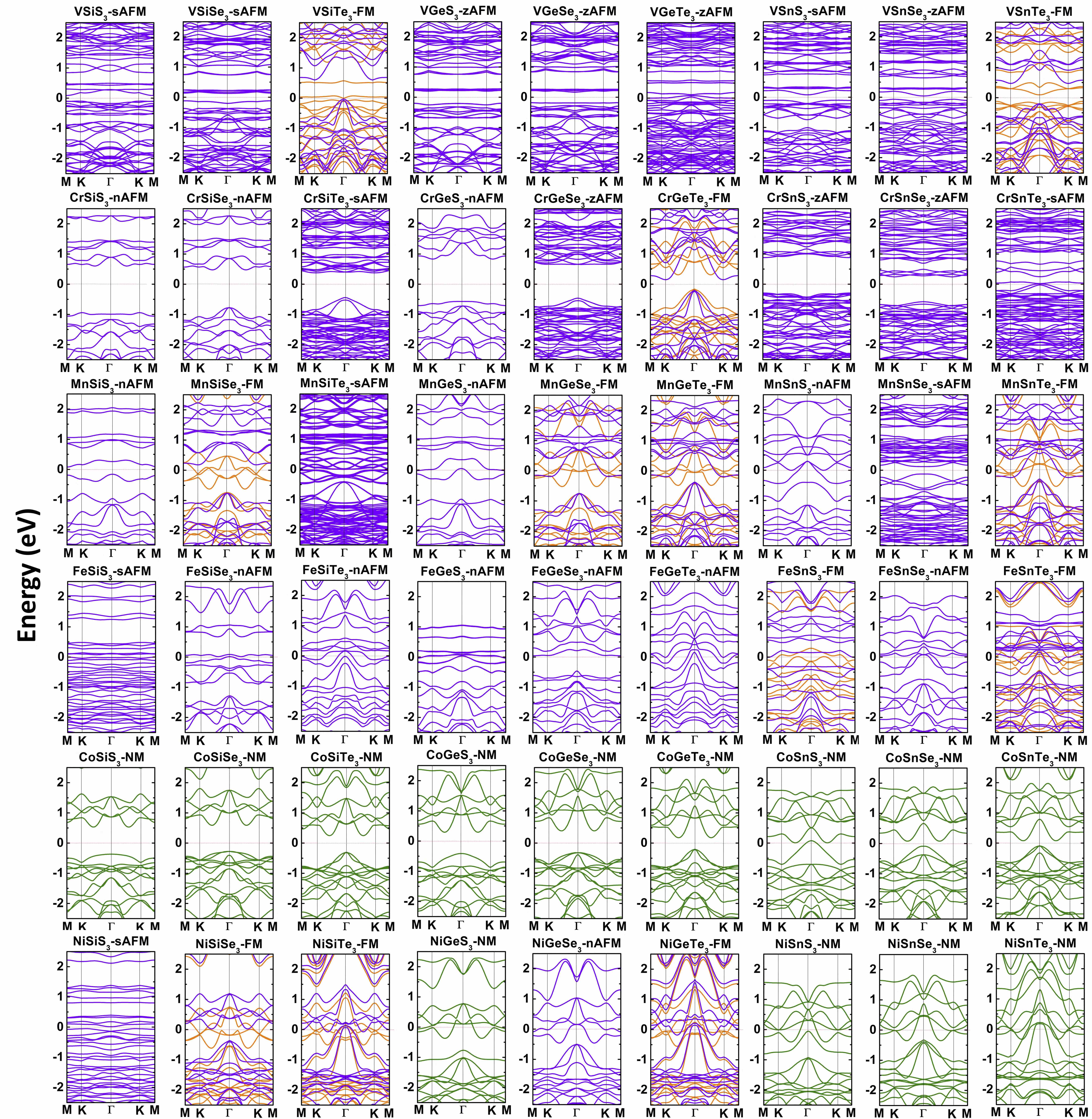}
\caption{(Color online) 
DFT-D2 band structures for single-layer MAX$_{3}$ compounds in their lowest-energy magnetic configurations for M = V, Cr, Mn, Fe, Co and Ni transition-metal atoms combined with A =  Si, Ge, Sn group IV and X = S, Se, Te chalcogen atoms. The plotted band structures were calculated using the triangular structural unit cell, except for the cases of sAFM and zAFM that have a larger periodicity in the magnetic structure.  There 
we used a triangular unit cell with doubled lattice constant. The bands are violet for AFM configurations, violet and orange for the up and down split spin bands in the FM configurations, and green for the NM phases.}
\label{BS}
\end{center}
\end{figure*}
\begin{sidewaysfigure*}[htbp!]
\centering
\textcolor{white}{\rule{6.4in}{3.6in}}
\includegraphics[width=24cm]{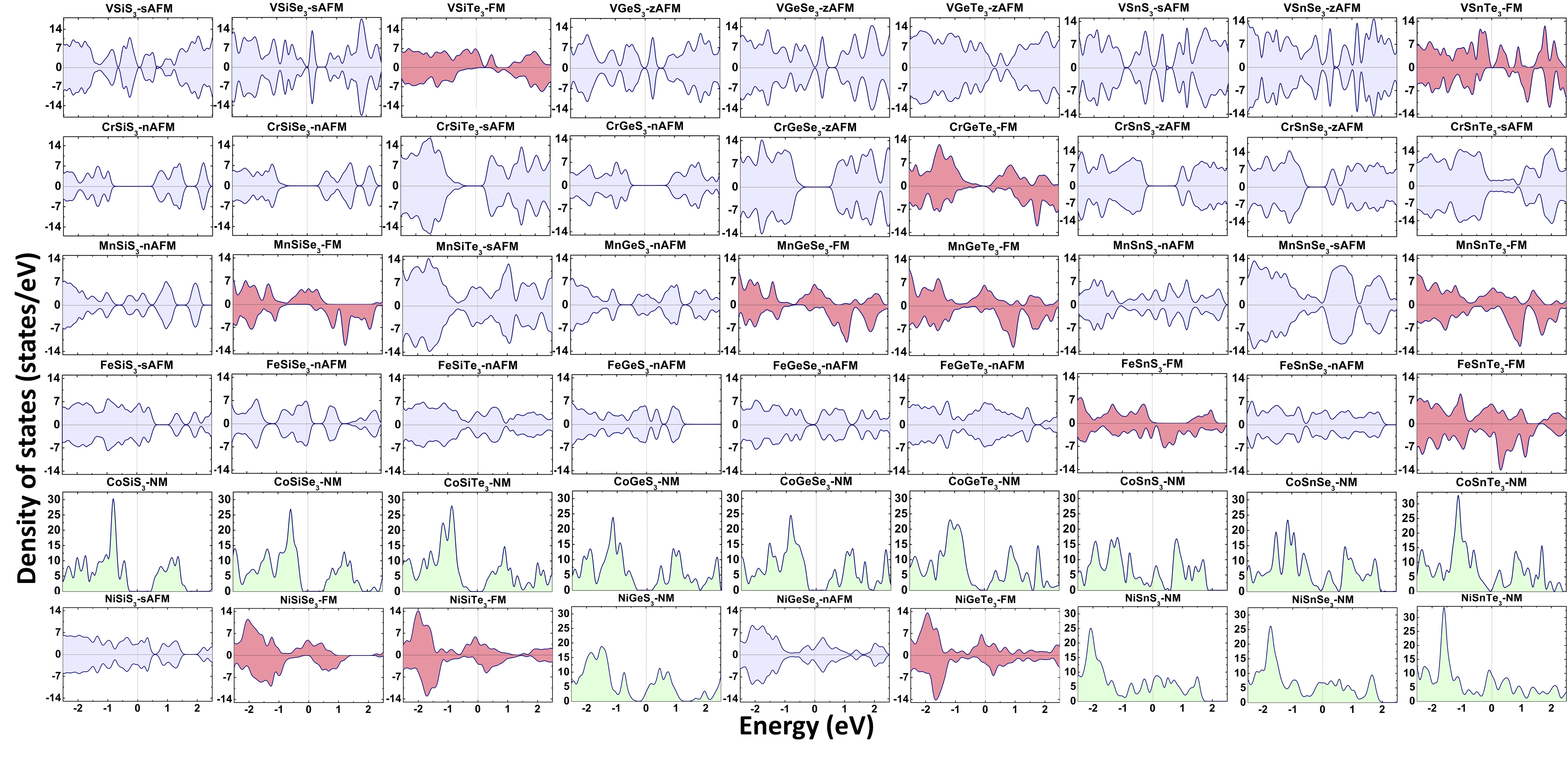}
\caption{
(Color online) 
The DFT-D2 density of states (DOS) for single-layer MAX$_{3}$ compounds in their lowest-energy magnetic configurations 
for M = V, Cr, Mn, Fe, Co and Ni transition-metal atoms with combination of A =  Si, Ge, Sn and X = S, Se, Te chalcogen atoms. 
Different colors, gray for AFM configurations, red for states in the FM configurations, and green for the NM phases are used to facilitate the 
classification of the expected  magnetic phases. 
Most ferromagnetic solutions are metallic except for VSnTe$_3$ and CrGeTe$_3$, 
while both gapped and metallic antiferromagnetic phases are common. 
}
\label{tri-DOS}
\end{sidewaysfigure*}
\begin{table*}[htb!]
\caption{
Magnetic moments in Bohr magneton $\mu_{\rm B}$ per metal atom for bulk and single 
layer magnetic MAX$_3$ structures, the three nearest neighbor exchange coupling strengths 
$J_i$ in meV implied by the Heisenberg model mapping, and Monte Carlo estimates of the 
single-layer critical temperatures based on the Ising limit of the classical spin model motivated by the 
perpendicular anisotorpy of these materials.  
The critical temperature is identified from the maximum in the derivative of magnetization with respect to temperature.  
The Ising limit approximation for the calculation of the T$_c$ is used as an upper bound for the Heisenberg models with magnetic anisotropy.
The {\it ab initio} calculations were performed within DFT-D2 and DFT-D2+U. 
In the cases of CoATe$_3$, CoASe$_3$ (A=Ge, Sn), and CoSnS$_3$ the U parameter values are 5, 6 and 7~eV, respectively. 
}
\begin{center}
\begin{tabular}{cc|cccc|ccc|c|cccc|ccc|c}\\ \hline \hline
                     &&   \multicolumn{8}{|c|}{DFT-D2} &   \multicolumn{8}{c}{DFT-D2+U}   \\ \hline
 MAX$_3$   & &{ FM }& {AFM}&{ zAFM }& {sAFM}&{$J_{1}$}&{$J_{2}$}&{$J_{3}$}&{T$_{\rm c}$}  &{ FM }& {AFM}&{ zAFM }& {sAFM}&{$J_{1}$}&{$J_{2}$}&{$J_{3}$}&{T$_{\rm c}$}\\ \hline
VSiS$_3$&&1.683&1.483&1.560&1.525&3.810&2.755&0.814&78 &1.903&1.880&1.903&1.874&2.138&0.230&0.310&125\\
VSiSe$_3$&&1.787&1.546&1.650&1.590&-3.441&0.926&1.231&123&2.466&2.612&2.607&2.409&1.763&0.013&-3.365&492\\
VSiTe$_3$&&1.932&1.768&1.890&1.895&-7.112&0.367&0.318&324&2.792&2.792&2.756&2.806&0.392&0.536&-0.372&117\\ \hline
CrSiS$_3$&&2.759&2.604&2.640&2.615&3.494&0.362&0.200&333&3.183&3.147&3.173&3.141&0.240&-0.0006&0.251&118\\
CrSiSe$_3$&&2.773&2.701&2.734&2.794&1.697&0.279&0.241&121&3.430&3.406&3.433&3.394&-0.713&0.027&0.425&128\\
CrSiTe$_3$&&2.851&2.794&3.076&3.031&-0.894&0.250&0.026&16&3.733&3.724&3.750&3.708&-1.266&-0.002&0.597&191\\ \hline
MnSiS$_3$&&3.612&3.423&3.503&3.553&0.705&-0.250&-0.480&122&4.232&4.468&4.453&4.234&-0.232&-0.230&-0.224&439\\
MnSiSe$_3$&&3.661&3.629&3.671&3.606&0.719&-0.382&-1.415&361&4.313&4.256&4.339&4.297&0.016&-0.389&-1.645&1110\\
MnSiTe$_3$&&3.741&3.719&3.765&3.660&0.471&-0.199&-1.019&260&4.410&4.443&4.463&4.396&-0.264&-0.171&-1.321&934\\ \hline
FeSiS$_3$&&0.983&1.025&1.013&1.012&2.738&1.701&-0.513&43&3.696&3.644&3.570&3.677&-0.184&0.325&1.729&541\\
FeSiSe$_3$&&1.082&1.149&1.143&1.121&2.402&1.190&1.052&22&3.557&3.510&3.560&3.547&-0.036&0.437&2.239&645\\
FeSiTe$_3$&&1.1603&1.247&1.230&1.190&16.394&0.654&-4.423&524&3.356&3.407&3.368&3.284&0.369&0.480&2.164&526\\ \hline
CoSiS$_3$&&-&-&-&-&-&-&-&-&2.181&2.220&2.264&2.205&1.224&-1.627&-3.479&677\\
CoSiSe$_3$&&-&-&-&-&-&-&-&-&1.200&1.241&1.254&1.202&-10.887&0.140&0.905&230\\
CoSiTe$_3$&&-&-&-&-&-&-&-&-&1.145&1.169&1.209&1.183&-1.759&2.994&1.977&110\\ \hline
NiSiS$_3$&&0.699&0.573&-&0.611&-&-&-&-&1.013&1.008&0.885&0.993&5.636&-10.646&-17.549&803\\
NiSiSe$_3$&&0.646&0.491&-&0.546&-&-&-&-&0.929&0.881&0.847&0.885&6.605&-13.366&-23.051&865\\
NiSiTe$_3$&&0.323&0.306&-&0.2741&-&-&-&-&0.739&0.632&0.503&0.649&1.848&-10.491&-19.897&396\\ \hline \hline
VGeS$_3$&&1.651&1.473&1.514&1.506&3.005&2.692&2.278&105&1.907&1.865&1.887&1.87&1.296&0.311&0.352&57\\
VGeSe$_3$&&1.660&1.532&1.575&1.653&1.128&1.712&1.628&109&2.083&2.602&2.109&2.599&-27.902&-0.174&10.083&904\\ 
VGeTe$_3$&&1.899&1.700&1.851&1.830&-1.969&0.962&-0.218&42&2.824&2.789&2.797&2.832&-0.918&-0.510&2.197&361\\ \hline
CrGeS$_3$&&2.675&2.597&2.624&2.610&1.776&0.324&0.221&135&3.10&3.130&3.155&3.137&-0.226&0.051&0.089&30\\
CrGeSe$_3$&&2.758&2.688&2.717&2.703&0.289&0.258&0.165&18&3.421&3.387&3.418&3.394&-0.939&0.115&0.176&41\\ 
CrGeTe$_3$&&3.078&2.774&2.803&3.024&-1.178&0.029&-0.243&237&3.725&3.709&3.733&3.708&-1.489&0.168&0.343&120\\ \hline
MnGeS$_3$&&2.798&2.529&2.654&2.587&1.569&0.008&-0.445&65&4.108&4.093&4.461&4.464&-0.680&-0.495&-0.210&847\\
MnGeSe$_3$&&2.980&2.879&2.887&2.852&-0.600&-0.134&-0.330&243&4.225&4.200&4.214&4.203&-0.152&-0.560&-0.909&1006\\ 
MnGeTe$_3$&&3.331&3.155&3.183&3.235&-0.409&-0.141&-0.193&229&4.364&4.3532&4.4458&4.435&0.228&-0.383&-0.666&583\\ \hline
FeGeS$_3$&&0.966&1.013&0.992&0.9940&0.252&-1.891&1.814&150&3.617&3.639&3.650&3.598&0.509&0.786&1.168&385\\
FeGeSe$_3$&&1.050&1.135&1.103&1.104&1.531&-0.145&1.532&98&3.549&3.576&3.506&3.530&0.319&0.548&1.058&325\\ 
FeGeTe$_3$&&1.141&1.230&1.224&1.210&26.517&-8.619&-16.300&562&3.356&3.409&3.367&3.425&0.439&0.833&1.015&308\\ \hline
CoGeS$_3$&&-&-&-&-&-&-&-&-&2.144&2.170&2.194&2.166&1.571&-2.863&-2.061&692\\
CoGeSe$_3$&&-&-&-&-&-&-&-&-&0.854&1.188&1.243&0.919&-6.818&5.778&18.977&556\\ 
CoGeTe$_3$&&-&-&-&-&-&-&-&-&1.077&1.036&1.189&0.864&-8.578&0.776&6.190&181\\ \hline
NiGeS$_3$&&-&-&-&-&-&-&-&-&0.981&0.973&0.897&0.949&0.637&-15.882&-8.083&943\\
NiGeSe$_3$&&0.499&0.518&0.197&-&-&-&-&-&0.903&0.856&0.807&0.822&1.056&-18.669&-7.102&847\\ 
NiSnTe$_3$&&0.088&-&-&-&-&-&-&-&0.086&0.1656&-&-&-&-&-&-\\ \hline \hline
VSnS$_3$&&1.576&1.302&1.312&1.509&-3.694&3.887&-0.477&75&1.871&1.831&1.852&1.842&0.753&0.644&0.603&41\\
VSnSe$_3$&&1.619&1.414&1.447&1.575&-4.667&2.334&0.471&98&2.104&1.948&1.983&2.0147&14.850&8.130&5.911&336\\
VSnTe$_3$&&1.602&1.466&1.660&1.611&-14.277&2.084&1.417&132&2.673&2.812&2.851&2.839&4.287&2.430&1.588&187\\ \hline
CrSnS$_3$&&2.6231&2.545&2.549&2.549&-1.775&1.077&1.015&266&3.0598&3.020&3.033&3.024&-1.810&0.165&0.230&116\\
CrSnSe$_3$&&2.692&2.622&2.704&2.731&-2.187&0.989&0.887&268&3.319&3.292&3.301&3.293&-1.838&0.159&0.198&165\\
CrSnTe$_3$&&2.894&2.895&2.825&2.937&1.023&1.667&-4.050&921&3.805&3.786&3.822&3.808&-1.489&0.447&0.02&25\\ \hline
MnSnS$_3$&&2.799&2.392&2.573&2.421&7.471&-0.592&-0.764&936&4.099&3.964&3.983&4.043&-0.499&-0.299&1.660&583\\
MnSnSe$_3$&&3.043&2.750&2.867&2.823&0.783&-0.192&-0.657&97&4.226&4.174&4.210&4.159&0.692&-0.354&-0.450&303\\
MnSnTe$_3$&&3.260&3.003&3.183&3.153&-0.091&-0.105&-0.365&169&4.458&4.456&4.473&4.464&-0.477&-0.593&-0.148&898\\ \hline
FeSnS$_3$&&0.929&0.938&0.924&0.947&-2.594&-0.706&-23.716&549&3.665&3.596&3.583&3.553&1.819&0.363&0.812&485\\
FeSnSe$_3$&&0.991&1.0142&0.986&1.013&-5.744&8.219&-4.317&100&3.446&3.444&3.466&3.449&0.149&-0.918&-0.150&475\\
FeSnTe$_3$&&1.088&1.193&1.114&1.170&0.327&-7.287&-0.967&462&3.117&3.043&2.932&3.111&2.910&0.245&-0.255&250\\ \hline
CoSnS$_3$&&-&-&-&-&-&-&-&-&0.914&0.593&0.913&2.042&-34.410&-12.166&26.680&386\\
CoSnSe$_3$&&-&-&-&-&-&-&-&-&1.024&0.663&1.046&1.044&-13.173&4.976&-1.686&43\\
CoSnTe$_3$&&-&-&-&-&-&-&-&-&0.855&0.638&1.117&1.090&-10.579&8.307&0.282&82\\ \hline
NiSnS$_3$&&&-&-&-&-&-&-&-&0.755&0.666&0.674&0.638&-1.554&-9.306&9.347&315\\ 
NiSnSe$_3$&&-&-&-&-&-&-&-&-&0.789&0.275&0.685&0.524&-29.469&5.580&-5.066&114\\ \hline \hline
\end{tabular}  \label{tab:magnetization}
\end{center}
\end{table*}

The magnetic ground-state and meta-stable magnetic configurations are obtained by identifying 
energy extremas via converged self-consistency initiated using initial conditions 
corresponding to N\'eel antiferromagnetic (nAFM), zigzag antiferromagnetic (zAFM), 
stripy aniferromagnetic (sAFM), ferromagnetic (FM), and nonmagnetic (NM) states.
The calculations we have carried out for single layer MAX$_3$ compounds 
confirm that the magnetic moments develop mainly 
at the metal atom sites, with significant net spin polarization induced on 
group IV and chalcogen atoms only in the ferromagnetic configuration case.
The late 3d transition metal elements Cr, Mn, Fe, Co and Ni stand out in the periodic table as elements that tend to order magnetically.   
The bonding arrangements of particular compounds can however enhance or suppress magnetism.
In 2D MAX$_3$ crystals, transition metal ions are contained within trichalcogenide  bipyramidal cages,
and have weak direct hybridization with other transition metal atoms.
The exchange interactions between the metal atoms are therefore mainly mediated
by indirect exchange through the intermediate chalcogen and A atoms.
Magnetic interactions can be extracted from {\it ab initio} electronic structure 
calculations by comparing the total energies of  antiferromagnetic, ferromagnetic, and nonmagnetic phases 
in V, Cr, Mn, Fe, Co, Ni based compounds as shown in Fig.~\ref{fig:groundstate}.
The total energy values in the data set used to construct the magnetic interaction 
model are gathered in 
Table.~I of the Supplemental Material\cite{SI}, where we find that 
magnetic phases are normally favored over the NM phase.  
Exceptions to this rule are that DFT-D2 predicts non-magnetic phases 
for CoAX$_3$, NiGeS$_3$ and NiSnX$_3$.  These compounds 
do develop magnetic solutions when we use a sufficiently large onsite repulsion parameter U within 
DFT-D2+U. 
We have generally used the U=4~eV value to assess the role of onsite Coulomb 
repulsion on the magnetic ground state energies, and
used larger values (U=5~eV for CoGeTe$_3$, CoSnTe$_3$, 
U=6~eV for CoGeSe$_3$, CoSnSe$_3$, 
and U=7~eV for CoSnS$_3$) to obtain magnetic solutions when they did not appear 
for U=4~eV.
Our DFT calculations predict that the Ni based trichalcogenides are magnetic 
only for some of the considered spin configurations. 
For instance NiGeS$_3$ and NiSnX$_3$ are non-magnetic within DFT-D2, 
and zAFM and sAFM phases are missing in NiSiX$_3$ and NiGeSe$_3$, 
and NiGeTe$_3$ has only ferromagnetic ordering. 
Within DFT-D2+U it is possible to obtain both FM and nAFM for NiGeTe$_3$, and FM only
for NiSnTe$_3$.

The magnetic anisotropy energy estimates that we obtained from non-collinear magnetization 
calculations are 616, 750 and 1166 $\mu$eV per formula unit for ferromagnetic order
in the compounds CrSiS$_3$, CrSiSe$_3$ and CrSiTe$_3$,
with magnetization favored perpendicular to the plane.
Assuming that in general these compounds will have a well defined anisotropy axis we make use of the 
Ising model for the spin-Hamiltonian to obtain an upper bound for the T$_c$ values.
The critical temperature estimate would be 
lower when we use a Heisenberg model with weak anisotropy. 

Using the fact that the magnetic moments are mostly concentrated at the metal atom sites
we can map the total energies to an effective classical spin Hamiltonian on a honeycomb lattice:
\begin{eqnarray}
H =  \sum_{\langle ij \rangle} J_{ij} \, \vec{S}_i \cdot \vec{S}_j = \frac{1}{2}  \sum_{ i \neq j } J_{ij} \, \vec{S}_i \cdot \vec{S}_j
\end{eqnarray}
where 
$J_{ij}$ are the exchange coupling parameters between two local spins,
$\vec{S}_i$ is the total spin magnetic moment of the atomic site $i$, 
and the prefactor 1/2 accounts for double-counting.
By calculating the three independent total energy differences between the four magnetic 
configurations~\cite{mpx3struct1,chaloupka,sivadas} illustrated in Fig.~\ref{fig:magneticorder}, 
namely the ferromagnetic (FM), N\'eel (nAFM),  zigzag AFM (zAFM), and stripy AFM (sAFM) configurations,
and assuming that the magnetic interactions are relatively short ranged, 
we can extract the nearest neighbor ($J_1$), second neighbor ($J_2$), 
and third neighbor ($J_3$) coupling constants:\cite{Bheema_PRB, sivadas}
\begin{eqnarray}
E_{\rm FM}-E_{\rm AFM} &=& 3 (J_1+J_3) \,  \vec{S}_{A} \cdot \vec{S}_{B}   \\
E_{\rm zAFM}-E_{\rm sAFM} &=& (J_1 - 3J_3) \,   \vec{S}_{A} \cdot \vec{S}_B  \\
E_{\rm FM}+E_{\rm AFM} &-& E_{\rm zAFM}-E_{\rm sAFM} = 8  J_2 \,  \vec{S}_{A} \cdot \vec{S}_A 
\end{eqnarray}
where $\vec{S}_{A/B}$ is the average spin magnetic moment on the honeycomb sublattice.
The average magnetic moment $S = \big{|}\vec{S}_{A/B}\big{|}$ at each lattice site obtained within DFT-D2 
and DFT-D2+U are listed in Table~\ref{tab:magnetization}.
For MAX$_3$ compunts where A=Si we get numerical values of magnetization for V, Cr, Mn, Fe close to 2, 3, 4, 1 Bohr magneton units while for A=Ge, Sn we get 2, 3, 3, 1.
The magnetization can be understood counting the 4s and 3d valence electrons in the metal atom used in the bonding with the anionic enclosure. 
For example, in the case of vanadium with three unpaired 3d electrons we expect that the three electrons required for the bonds originate from two 4s electrons and one 3d electron, 
which leaves two upaired 3d electrons responsible fo the 2$\mu_B$.
The magnetization increases as we move to the right in the periodic table until it saturates for Mn with the maximum of five unpaired 3d electrons.
We notice that the sudden drop in magnetization for Fe down to one Bohr magneton may be attributed to changes in the 4s and 3d energy level ordering 
such that three out of the four unpaired 3d valence electrons are used for the bonds with the enclosure.

Single-layer magnetic ordering temperatures T$_{\rm c}$ were estimated by running Monte Carlo simulations of the three-coupling-constant Ising models
using the Metropolis algorithm in lattice sizes of up to $N$=32$\times$64 with periodic boundary conditions~\cite{metropolis,newman,landau}
calculating the heat capacity $C = k \beta^2 \left( \langle E^2 \rangle - \langle E \rangle^2 \right)$ as a function of temperature and identified its diverging point as the Neel and Curie temperatures.
This value can be considered as an upper bound for a Heisenberg model with strong anisotropy. 
See the Supplemental Material\cite{SI} for plots of representative results 
for the temperature dependent heat capacity.
The calculated average values of the magnetic moments 
vary widely from compound to compound (See Table~\ref{tab:magnetization}), 
while the magnitudes of the magnetic moments at the metal atoms generally have relatively 
small differences (between 3\%$\sim$10\%) between different magnetic configurations of the same 
compound.  The largest variations within a compound were found in FeSnX$_3$ and NiSnX$_3$. 
Within the DFT-D2 approximation we find that the magnetic moments in 2D MAX$_3$ develop 
almost entirely at the metal atom sites while the use of on-site U introduces an enhancement of the spin 
polarization at non-metal atom sites.  Variation of moment from configuration to configuration
within a compound is an indication that the system is less accurately described by a 
local moment model.

\subsection{Band structure and density of states}
Understanding the electronic properties is an essential stepping stone 
to seek spintronics device integrations based on MAX$_3$ magnetic 2D materials.
Specifically, it is desirable to understand how the electronic structure depends on the type of magnetic phase
in order to seek ways to couple charge and spin degrees of freedom.
\begin{table*}[htb!]
\caption{Band gaps and electronic nature of MAX$_3$ compounds. The band gaps are listed in eV energy units and their values
and the magnetic configuration depend substantially on the exchange-correlation approximation employed within DFT-D2 and DFT-D2+U (=4~eV). 
For CoATe$_3$, CoASe$_3$ (A=Ge,Sn), CoSnS$_3$ the U parameter values chosen are 5, 6 and 7~eV respectively.
Different calculation methods have been indicated by I:DFT-D2 and II:DFT-D2+U. 
The ground states for selected method are represented in boldface type and blue color (M:Metal; SM:Semi-Metal; HM:Half-Metal). 
}
\begin{center}
\begin{tabular}{cc|ccccc|ccccc|ccccc}\\ \hline \hline
A&& \multicolumn{5}{c|}{Si} & \multicolumn{5}{c|}{Ge} & \multicolumn{5}{c}{Sn} \\ \hline
MAX$_3$&Method&{NM}&{FM}&{nAFM}&{zAFM}&{sAFM}&{NM}&{FM}&{nAFM}&{zAFM}&{sAFM}&{NM}&{FM}&{nAFM}&{zAFM}&{sAFM}\\ \hline
VAS$_3$&I &M &HM &SM &M &{\color{blue} \bf 0.106} &M &HM &SM &{\color{blue} \bf 0.533} &SM &M &M &M &{\color{blue} \bf 0.180} &M \\
&II&0.541 &1.227 &{\color{blue} \bf 0.994} &1.227&1.200& 0.975 &1.371 &{\color{blue} \bf 1.347} &1.371&1.335&0.252 &1.155 &1.226 &{\color{blue} \bf 1.017}&1.263 \\ \hline
VASe$_3$&I &M &HM &SM &{\color{blue} \bf 0.255} &M &M &HM &SM &{\color{blue} \bf 0.350} &SM &M &M &M &{\color{blue} \bf1.030} &M \\
&II&0.504 &M &SM &SM&{\color{blue} \bf 0.251}&0.325 &0.974 &0.215 &{\color{blue} \bf 1.555}&0.143 &SM &M & 0.938&{\color{blue} \bf 1.497}&0.829 \\ \hline
VATe$_3$&I &M &{\color{blue} \bf HM} &SM &M &M &M &HM &M &{\color{blue} \bf M} &M &M &{\color{blue} \bf 0.216} &M &M &M \\ 
&II &0.289 &M &M&M &{\color{blue} \bf M} & M &M &{\color{blue} \bf M}&M&M &SM &M &M & {\color{blue} \bf M}&M \\ \hline
CrAS$_3$& I &M &0.785 &{\color{blue} \bf 1.544}&1.082&1.001 &M &0.513 &{\color{blue} \bf 1.405}&0.947&0.974 &M &0.324 &0.812&{\color{blue} \bf 1.173}&0.757 \\
&II &M &0.613 &{\color{blue} \bf 1.176}& 0.865&1.047 &1.390 &0.794 &0.938&{\color{blue} \bf 1.311}&0.793 &0.613 &{\color{blue} \bf 0.544} &1.154&1.047&1.010 \\ \hline
CrASe$_3$& I &M & 0.731 &{\color{blue} \bf 1.402}&1.010&0.839 &M &0.568 &0.812&{\color{blue} \bf 1.121}&0.703 &M & 0.107 &0.460&{\color{blue} \bf 0.823}&0.406 \\ 
&II &M &0.214 &0.558&{\color{blue} \bf 0.937}&0.454 &M &0.252 &0.540&{\color{blue} \bf 0.860}&0.427 &M &{\color{blue} \bf  0.465} &0.468&0.505&0.432 \\ \hline
CrATe$_3$& I &M &0.405&0.541&{\color{blue} \bf 0.817}& 0.378 &M &{\color{blue} \bf 1.040} &0.324&0.432&0.270 &M &M &M &M &{\color{blue} \bf M} \\ 
&II &M &SM &SM&{\color{blue} \bf 0.623}&SM &SM &SM &0.180&{\color{blue} \bf 0.527}&0.107 &M &HM &SM&{\color{blue} \bf 0.367}&SM \\ \hline 
MnAS$_3$& I &M &M &{\color{blue} \bf 0.267} &SM &M &M &M &{\color{blue} \bf M} &SM &SM &M &M &{\color{blue} \bf M} &M &M \\
& II&0.757 &{\color{blue} \bf HM }&0.469 &0.469&M & 0.694&{\color{blue} \bf HM }&M&0.710&SM &M &HM &{\color{blue} \bf 0.555}&M&M \\ \hline 
MnASe$_3$&I &M &{\color{blue} \bf HM} &M &M &M &M &{\color{blue} \bf M} &M &SM &SM &M &M &M &M &{\color{blue} \bf0.245} \\
&II &0.077 &{\color{blue} \bf HM} &M &M&M&0.999 &{\color{blue} \bf HM} &M &M&M &0.611 &HM &{\color{blue} \bf M}&M&M \\ \hline
MnATe$_3$&I &M &M &M &M &{\color{blue} \bf M} &M &{\color{blue} \bf M} &M &M &M &M &{\color{blue} \bf M} &M &M &M \\
&II &0.650 &{\color{blue} \bf HM} &M &M&M&0.599 &{\color{blue} \bf HM} &M &M&M &0.721 &{\color{blue} \bf HM} &M&M&M \\ \hline 
FeAS$_3$&I &M &0.541 &0.622 &0.486&{\color{blue} \bf M} &M &0.649 &{\color{blue} \bf M} &0.785&0.730&M &{\color{blue} \bf HM} &0.378&0.487&0.510 \\
&II &0.751 &HM & SM&{\color{blue} \bf 0.730}&SM &0.799 &HM &SM&{\color{blue} \bf 0.730}&SM &M &HM & {\color{blue} \bf 0.302}&0.287&SM \\ \hline
FeASe$_3$&I &M &0.378 &{\color{blue} \bf M}&0.703&0.514 &M &0.432 &{\color{blue} \bf 0.507}&0.622&0.486 &M &0.334 &{\color{blue} \bf M}&0.433&0.406 \\
& II&0.173 &M &SM&{\color{blue} \bf 0.655}&M &0.099 &M &SM&{\color{blue} \bf 0.567}&M &M &{\color{blue} \bf HM} &0.100&SM&M \\ \hline
FeATe$_3$&I &M &0.243 &{\color{blue} \bf M}&0.487&0.378 &M &0.243 &{\color{blue} \bf M}&0.433&0.270 &M &{\color{blue} \bf M} &0.189&0.108&0.054 \\ 
&II &M &M &SM&{\color{blue} \bf 0.347}&M &M &M &M&{\color{blue} \bf 0.332}&M &M &M &M&M&{\color{blue} \bf M} \\ \hline 
CoAS$_3$&I &{\color{blue} \bf 0.900} &- &-&-&- &{\color{blue} \bf 0.706} &- &-&-&- &{\color{blue} \bf M} &- &-&-&- \\
&II &{\color{blue} \bf 1.260} &HM &SM&M& M & {\color{blue} \bf 1.033}& HM &SM&M&M &1.046 &{\color{blue} \bf HM} &0.902&0.505& M \\ \hline
CoASe$_3$&I &{\color{blue} \bf 0.865} &- &-&-&- &{\color{blue} \bf 0.728} &- &-&-&- &{\color{blue} \bf 0.167} &- &-&-&- \\
&II &0.650 &{\color{blue} \bf HM} &SM&M& M &0.599 &HM &SM&{\color{blue} \bf 0.860}&SM &0.577 &0.180 &0.454&{\color{blue} \bf 0.466}&SM \\ \hline
CoATe$_3$&I &{\color{blue} \bf 0.864} &- &-&-&- &{\color{blue} \bf 0.305} &- &-&-&- &{\color{blue} \bf 0.172 }&- &-&-&- \\
&II &0.288 &HM &SM&{\color{blue} \bf0.209}&M &SM &HM &SM&{\color{blue} \bf 0.166}&SM &SM &M &SM&{\color{blue} \bf M}&M \\ \hline
NiAS$_3$&I &M &M &M &- &{\color{blue} \bf M} &{\color{blue} \bf M} &- &-&-&- &{\color{blue} \bf M} &- &-&-&- \\
&II &M &{\color{blue} \bf HM}& M&M& M &M &{\color{blue} \bf HM} &M&M&M &M &M &{\color{blue} \bf M}&SM&M \\ \hline
NiASe$_3$&I &M &{\color{blue} \bf HM} &M &- &M &M &M &{\color{blue} \bf M} &M&- &{\color{blue} \bf M} &- &-&-&- \\
&II &M &{\color{blue} \bf HM}&M&M& M & M&{\color{blue} \bf HM} &M&M&M &M &{\color{blue} \bf HM} &M&M&M \\ \hline
NiATe$_3$&I &M &{\color{blue} \bf M} &M &- &M &M &{\color{blue} \bf M} &-&-&- &{\color{blue} \bf M} &- &-&-&- \\ 
&II & M&{\color{blue} \bf M}&M&M&M &M &{\color{blue} \bf M}& M &-&- &M &{\color{blue} \bf M} &-&-&- \\ 
\hline \hline
\end{tabular} \label{tab:gaps}
\end{center}
\end{table*}
It is found that the MAX$_3$ class of materials includes almost all
of the behaviors studied in current spintronics research, including 
both antiferromagnets and ferromagnets, and a variety of electrical properties including metals, semi-metals, half-metals and semiconductors.
We have classified as semi-metallic those states with vanishingly small gaps or small density of states (DOS) near the Fermi energy
from inspection of the electronic structure. 
The electronic band structures corresponding to the ground state configurations are represented in Fig.~\ref{BS}
and the respective DOS are in Fig.~\ref{tri-DOS}.
The band structures are plotted using a triangular unit cell around one of the $K$ valleys, at times doubling the cell size to allow for longer period magnetic configurations.
The DOS for all magnetic (FM, nAFM, zAFM and sAFM) and non-magnetic phases within DFT-D2 and DFT-D2+U can be found in the Supplemental Material\cite{SI}. 
The analysis of the orbital projected partial density of states in the Supplemental Material\cite{SI} for the MAX$_3$ 
compounds reveals that the conduction band edges have an important contribution from the $d$ orbitals of metal atoms, 
as well as $s$ and $p$ orbitals of the A atom for the valence band edges. 
From the difference in the associated density of states between DFT+D2 and DFT+D2+U we notice the strong sensitivity of the electronic structure to the choice of electron-electron interaction model.

As expected, most of the AFM cofigurations are found to be semiconductors while semi-metallic and metallic solutions are also 
found for select spin configurations in V and Cr tellurides, in two instances of Mn sulfides, and in several Fe based compounds, see Fig.~\ref{BS}.
As a general trend, we notice that both AFM as well as NM band gaps reduce when the chalcogen's atomic number increases from S to Te. 
The FM configurations are generally metallic, with half metallic solutions for VSiTe$_3$,  MnSiSe$_3$,  FeSnS$_3$,  NiSiSe$_3$, 
and are semiconducting for CrGeTe$_3$, FeSnS$_3$, FeSnTe$_3$. 
We notice that the addition of U switches some of the metallic FM solutions into half-metals, and it leads to semiconducting FM solutions for CrSnS$_3$, CrSnSe$_3$ and CoSnS$_3$.
Most of the Co based compounds predict NM states with a semiconducting gap,
while the few NM states of Ni based compounds are found to be metallic. 

We notice that the magnitude of the band gaps in AFM, FM and NM states do not experience notable changes between 
DFT-D2 and DFT-D2+U, as illustrated in Table~\ref{tab:gaps}.
However, the corresponding density of states (DOS), plotted in Fig.~\ref{tri-DOS}, 
shows the strong influence on the ground-state electronic structure when Coulomb correlations are
included. This suggests that the physics of MAX$_3$ compounds can be dominated by correlation effects
and modeling will be most successful when we rely on effective models that feed from experimental input or
high level {\em ab initio} calculations. The orbital content of the valence and conduction band edges
that are most relevant for studies of carrier-density dependent magnetic properties
can be extracted from the orbital projected partial density of states (PDOS).
Depending on the specific material composition and the magnetic configuration,
the valence and conduction band edge orbitals can be dominated 
by metal, non-metal or chalcogen atoms. 

\section{Tunability of magnetic properties}
\label{sec:tunability}
As we reported in the case of MPX$_3$,\cite{Bheema_PRB} the two dimensional magnetic materials are of interest 
primarily because of the prospect that 
their properties might be more effectively altered by tuning parameters controllable {\it in situ}. 
Two potentially important control knobs that can be exploited experimentally
in two-dimensional-material based nano-devices are 
carrier-density and strain.  The dependence of magnetic properties on 
carrier density is particularly interesting because it provides a convenient route
for electrical manipulation of magnetic properties in electronic devices.
In this section we explore the possibility of tailoring the electronic and magnetic properties of MAX$_3$ ultrathin layers
by adjusting carrier density or by subjecting the MAX$_3$ layers to external strains.

\begin{figure*}[htb!]
\includegraphics[width=20cm]{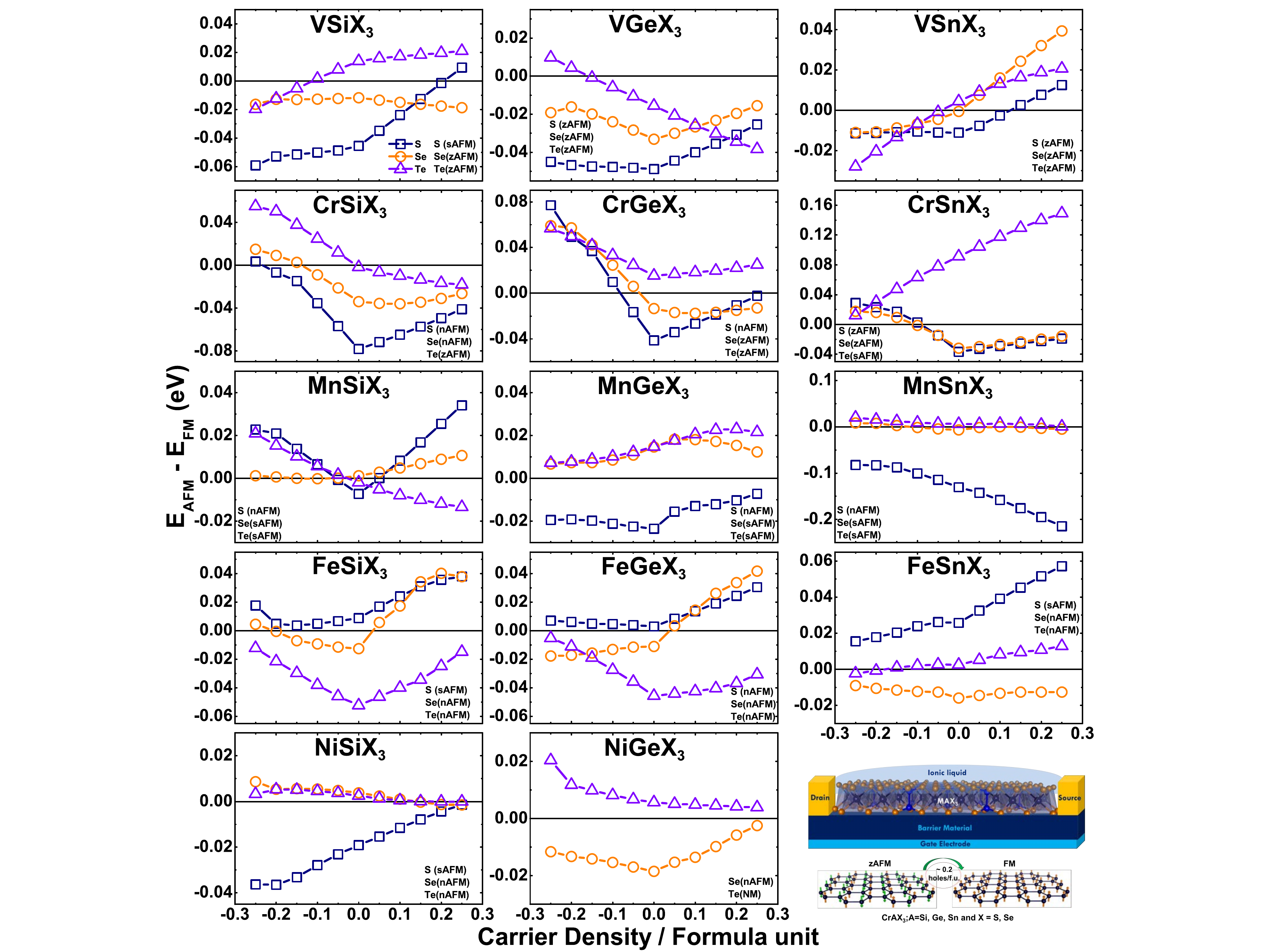}
\caption{(Color online) 
Carrier density dependent total energy differences per MAX$_3$ formula unit between the AFM and FM phases 
of V, Cr, Mn, Fe, Ni based single layer trichalcogenides obtained within DFT-D2.  
The AFM ground-states favored near charge neutrality can often be switched to FM phases either for electron or hole doping 
for sufficiently large carrier densities in V, Mn, Ni, and Fe based compounds.
Carrier densities of up to a few $ {\sim10^{14} } $ electrons per ${\rm cm}^2$ should be in principle accessible through 
ionic liquid or gel gating.
A carrier density of 0.1 electrons per MAX$_3$ formula unit corresponds to $\sim$${6\times 10^{13}}$ electrons per cm$^2$ 
when the distance between the metal atoms is $\sim$6$\angstrom$.}
\label{carrier_magnetism}
\end{figure*}

\subsection{Field-effect modification of magnetic properties}
The possibility of modifying the magnetic properties of a material simply by applying a gate voltage offers advantages over 
magnetic-field mediated information writing in magnetic media such as the higher density storage and enhanced information access speed.  
Electric field control of magnetic order has been demonstrated in ferromagnetic semiconductors and metal films through carrier 
density or Fermi level dependent variation of magnetic exchange or magnetic anisotropy.~\cite{ohno1,ohno2,endo}
The gate voltage control of magnetism in 2D materials would allow control over magnetically stored information
at negligible energy cost. 

\begin{figure}[htb!]
\begin{center}
\includegraphics[width=\columnwidth]{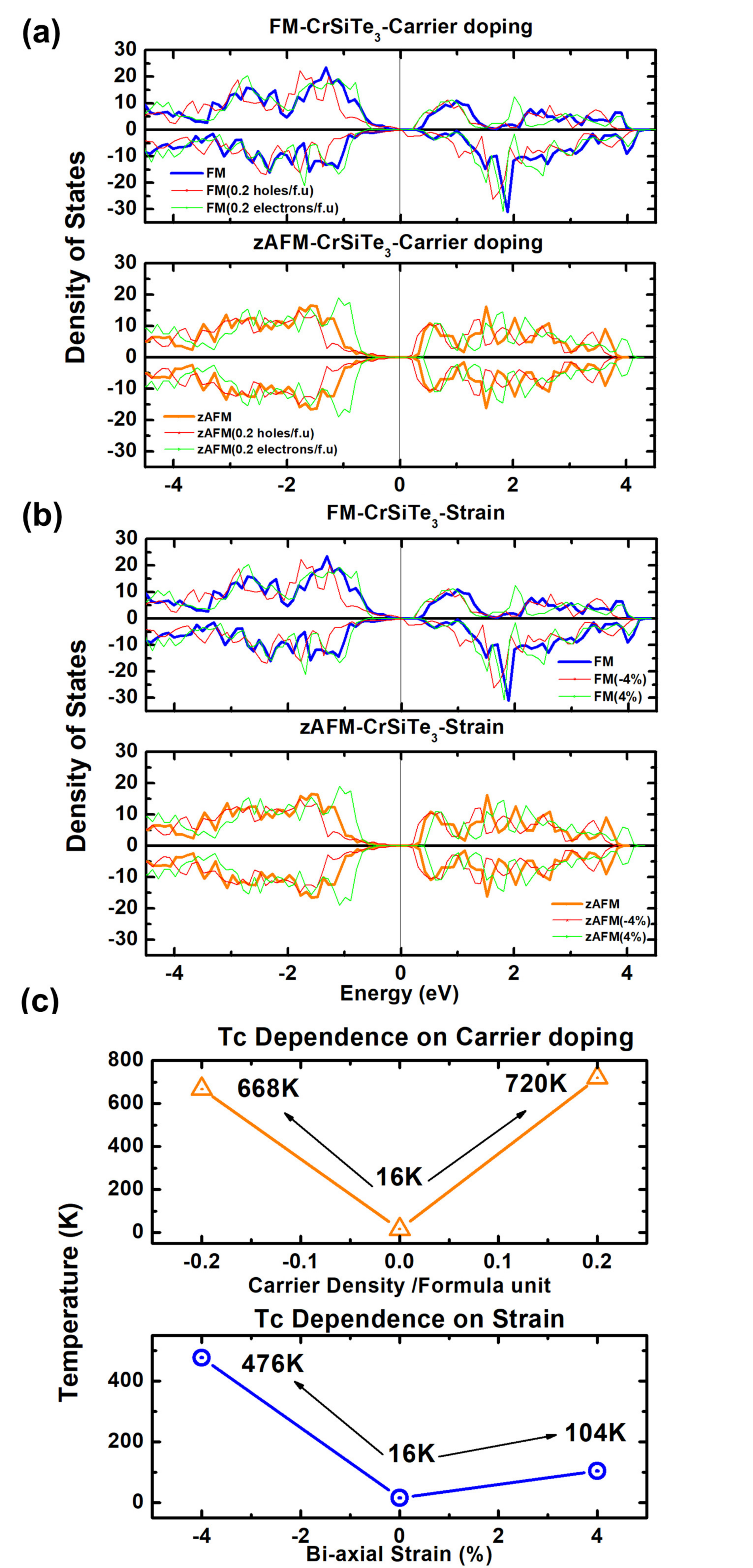}
\caption{
(Color online) The comparison of Density of states (DOS) and Projected DOS in the presence of 
(a) carrier doping of $-$0.2~(holes per formula unit) and 0.2~(electrons per formula unit), and 
(b) subject to biaxial strains of -4{\%} (compressive) and 4{\%} (expansive) in the FM and zAFM states in CrSiTe$_3$ for FM and zAFM phases.
(c) We observe substantial enhancements in the T$_c$ values as a function of carrier density and strains.}
\label{carrier-str-dos-tc}
\end{center}
\end{figure}

A schematic illustration for a field effect transistor device where magnetic order is modified
through a carrier inducing backgate is shown in Fig.~\ref{carrier_magnetism}, where we also 
summarize the theoretically predicted trends for the competition 
between AFM and FM states for 2D MAX$_3$ compounds that we have considered.  
In our calculations we have obtained the variation of the total energy differences between magnetic configurations as a function of 
carrier density neglecting the possible role of charge polarization within a MAX$_3$ layer induced by the external fields. 
We find that when the ground state is in the AFM phase, 
generally transitions to FM phases can be achieved by adding sufficiently large electron or hole carrier densities. 
The origin of this trend can be understood based on energetic considerations 
when FM phases are gapless or have smaller gaps than the AFM phases.~\cite{Bheema_PRB} 
If we denote by $\Delta E_{gap} = E_{\rm AFM}^{gap} - E_{\rm FM}^{gap}$ the difference between the gap in the AFM and
the FM phases, it follows that the energy difference per area unit between AFM and FM 
phases $\delta E \equiv E_{\rm AFM} -E_{\rm FM}$ is given at low carrier densities by 
\begin{eqnarray}
  \delta E(\pm \delta n) &=& \delta E_0  + {  ( \Delta E_{gap}/2 \pm  \delta \mu ) \, (\pm \delta n)   } 
\end{eqnarray} 
where $+\delta n$ is the carrier density of $n$-type samples, and $-\delta n$ is the carrier density of $p$-type samples, 
$\delta E_0$ is the energy difference per area unit between AFM and FM phases in neutral 
MAX$_3$ sheets, and $\delta \mu$ is the difference between the mid-gap energy of the 
AFM semiconductors and the chemical potential of the ferromagnetic metals or semiconductors.  
Introducing $n$-carriers is most effective in driving a transition from AFM to 
FM phases when $\delta \mu$ is positive, while introducing $p$-carriers is 
most effective when $\delta \mu$ is negative.  
%
%

The total DOS corresponding to FM and AFM phases in the presence of carrier doping
is illustrated for CrSiTe$_3$ in Fig.~\ref{carrier-str-dos-tc}(a) as a specific example.   
A detailed breakdown of the projected density of states at each atomic site 
can be found in the Supplemental Material~\cite{SI}.

Because the energy difference per formula unit between FM and AFM phases is 
much smaller than the energy gap, a transition between them
can be driven by carrier density changes per formula unit that are much smaller than one, 
especially so when $\delta \mu$ plays a favorable role.  
In particular we see in Fig.~\ref{carrier_magnetism} that a 
transition between FM and AFM phases are predicted in 
CrSiTe$_3$, MnSiX$_3$, CrGeSe$_3$ at electron carrier densities and 
MnSiS$_3$, FeSiSe$_3$, FeGeSe$_3$, VSnSe$_3$ at hole densities as small as $\sim$0.05 electrons per formula unit
which correspond to carrier densities on the order of ${10^{13}} {\rm cm}^{-2}$.
Our calculations show that the FM solution is the preferred stable magnetic configuration 
in almost all cases when the system is subject to large electron or hole densities within the 
range of a few times ${\pm 10^{14}}$ cm$^{-2}$.
Even then, cases like vanadium based VSiSe$_3$, VGeS$_3$, VGeSe$_3$, 
or iron based FeSiTe$_3$, FeSnS$_3$, \& FeSnSe$_3$, 
or manganese based MnGeX$_3$, MnSnS$_3$,  or NiGeX$_3$, CrSnTe$_3$,  
have not shown any transition within the selected range of electron or hole carrier density. 
Carrier density changes of this magnitude can be achieved by ionic liquid or gel gating, or through interfaces with ferroelectric materials.  
Since this size of carrier density is sufficient to completely change the character of the magnetic order, 
we can expect substantial changes in magnetic energy landscapes and their stability at much smaller carrier densities.   

Our calculations therefore motivate efforts to find materials which can be used to establish 
good electrical contacts to MAX$_3$ compounds to facilitate either $n$ or $p$ carrier doping 
by aligning their fermi levels towards the conduction or valence bands.
For compounds whose ground-states are FM at charge neutrality 
we find that the transitions to the AFM phase can be achieved 
for $n$-doping in VSiTe$_3$, VSnTe$_3$, MnSiSe$_3$, \& FeSnTe$_3$
and for $p$-doping for NiSiSe$_3$, NiSTe$_3$, \& MnSnTe$_3$.

\begin{figure*}[htb!]
\includegraphics[width=20cm]{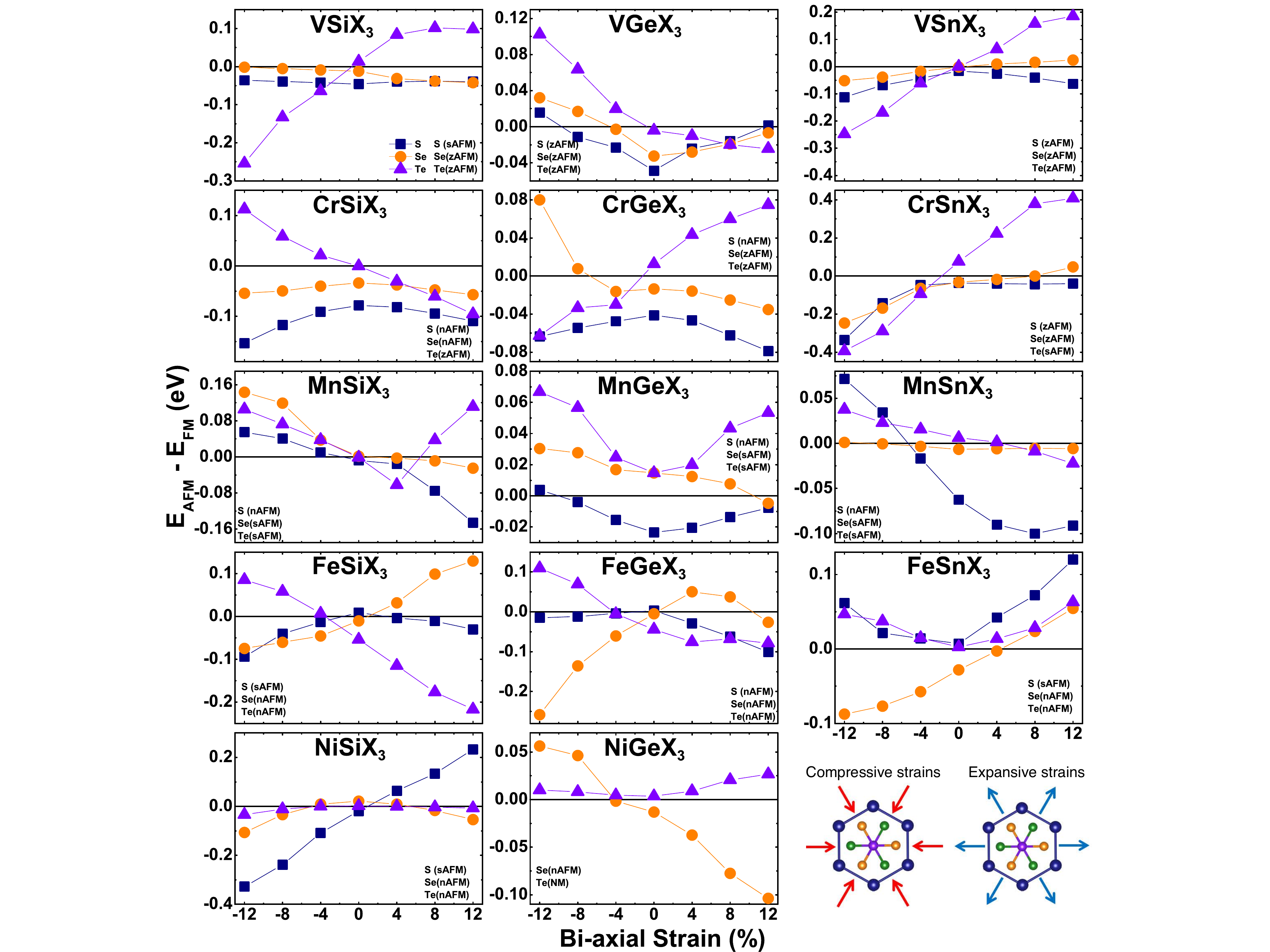}
\caption{(Color online) 
Influence of in-plane strain on the magnetic configurations of selected MAX$_3$ compounds. 
Magnetic phase transitions are introduced by in-plane biaxial compressive and expansive strains for several magnetic 
MAX$_3$ compounds at zero carrier density. We notice that the magnetic phase energy differences in some compounds are especially susceptible 
to the strains suggesting that large variations in T$_c$ values may be achievable by strain engineering.}
\label{fig:strain}
\end{figure*}

\subsection{Strain-tunable magnetic properties}
The flexible membrane-like behavior of 2D-materials can be used to tailor their electronic structure by means of strains.
Examples of strain induced electronic structure modification in 2D materials discussed in the 
literature include the observation of Landau-level like density of states near high curvature 
graphene bubbles \cite{guineabubbles},
or the commensuration moir\'e strains that open up a band gap at the primary Dirac point 
in nearly aligned graphene on hexagonal boron nitride~\cite{jarillogap,woods,origingap}.
Noting that strains can play an important role in configuring the electronic structure
we calculate the total energies for different magnetic phases in 2D MAX$_3$ materials
in the presence of expansive or compressive in-plane biaxial strains that we model by
uniformly scaling the rectangular unit cell described in Fig.~\ref{fig:cellstructure}. 
The uniform biaxial strains lead to modifications in the magnetic phase 
energy differences $E_{\rm AFM} - E_{\rm FM}$ that can trigger phase transitions 
for strains as small as 2-4\% in certain cases, while much larger strain fields are required in general.
In the following we list the five different types of strain-induced effects expected in charge neutral MAX$_3$ compounds:

\begin{itemize}
\item { No phase change (expansion and compression):} 
The ground-states are not altered in the presence of strains for 
vanadium based VSiS$_3$ (sAFM), VSnS$_3$ (zAFM), 
chromium based CrSiS$_3$ (nAFM), CrSiSe$_3$ (nAFM), CrGeS$_3$ (nAFM), CrSnS$_3$ (zAFM),
manganese based MnGeTe$_3$ (FM), nickel based NiGeTe$_3$ (FM),  
and iron based FeSnS$_3$ (FM), FeSnTe$_3$ (FM).

\item { AFM to FM (compression):} 
For compressive strains we find transitions for
vanadium based VSiSe$_3$ (4\%), VGeS$_3$ ($\sim$1\%), VGeSe$_3$ (4\%), VGeTe$_3$ (9\%),
chromium based CrSiTe$_3$ ($\sim$1\%), VGeTe$_3$ (9\%),
manganese based MnSiS$_3$ (2\%), MnSiTe$_3$ ($\sim$1\%), MnGeS$_3$(9\%), MnSnS$_3$ (4\%), MnSnSe$_3$ (8\%),
nickel based NiGeSe$_3$ (4\%), and
iron based FeSiTe$_3$ (4\%), FeGeTe$_3$ (4\%).

\item { AFM to FM (expansion):}
Conversely for expansive strains we find transitions 
in vanadium based VGeS$_3$ ($\sim$12\%), VSnSe$_3$ ($\sim$2\%)
chromium based CrSiTe$_3$ ($\sim$2\%), CrSnSe$_3$ (8\%),
manganese based MnSiTe$_3$ ($\sim$6\%),
and iron based FeSnSe$_3$ (4\%).

\item { FM to AFM (compression):} 
Transitions are seen for compressive strains (\%) in vanadium based VSiTe$_3$ ($\sim$1\%), 
VSnTe$_3$ ($\sim$1\%), nickel based NiSiSe$_3$ (4\%), NiSiTe$_3$ (4\%), 
and chromium based CrGeTe$_3$ ($\sim$2\%), CrSnTe$_3$ ($\sim$2\%).

\item { FM to AFM (expansion):} Transitions for expansive strains (\%) are found for manganese based MnSiSe$_3$ ($\sim$1\%), MnSnTe$_3$ (4\%), iron based FeSiS$_3$ (4\%), and nickel based NiSiSe$_3$ (4\%), NiSiTe$_3$ (4\%).
\end{itemize}

An example about the DOS evolution as a function of strain is shown in Fig.~\ref{carrier-str-dos-tc} (b)
for charge neutral CrSiTe$_3$ monolayer subjected to $-$4{\%} (compressive) and 
4{\%} (expansive) strains in the FM and zAFM phases.
The expansion strains are found to have a small effect in both the FM and zAFM phases of CrSiTe$_3$ 
but compressive strains of 4\% lead to a closure of the 
FM-CrSiTe$_3$ band gap turning it into a semi-metal with finite density of states at the Fermi level.
As mentioned earlier, from the projected density of states analysis in the Supplemental Material \cite{SI}
we can observe a relatively large content of Cr-$d$ and Si-$s, p$ orbitals at the bands near the Fermi energy.

The Fig.~\ref{carrier-str-dos-tc}(c) shows that the T$_c$ can be substantially enhanced when the device is subject
to carrier density variations or to strains. 
For monolayer CrSiTe$_3$ we showed 
that the T$_c$ increased from 16K to almost 700K
for carrier densities of 0.2 electrons (holes) per CrSiTe$_3$, while in the presence of strains
they enhanced to 476K  (4\% compression) and 104K (4\% expansion).

\section{Summary and discussion}
\label{sec:summary}
In this paper we have carried out an {\em ab initio} study of the MAX$_3$ transition metal 
trichalcogenide class of two-dimensional materials considering different 
combinations of 3d metal (M = V, Cr, Mn, Fe, Co, Ni), group IV (A = Si, Ge, Sn) and chalcogen (X = S, Se, Te)
atoms in an effort to make an exhaustive search for magnetic 2D materials useful for spintronics applications.
Our calculations suggest that magnetic phases are common in the single-layer limit of these 
van der Waals materials, and that the configuration of the magnetic phase depends sensitively 
on the transition metal/chalcogen element combination. 

We find that semiconducting antiferromagnetic (AFM) phases are the most common ground state configurations
and appear in N\'eel, zigzag, or stripy configurations.
The ferromagnetic (FM) phases are predominantly metallic although semiconducting band structures are found for several compounds, while 
the non-magnetic (NM) phases exist for both metallic and semiconducting states. 
In compounds with larger chalcogen atoms we have relatively smaller band gaps in 
the AFM phases and smaller ground state energy differences between the FM and AFM phases.
Compounds such as CoAX$_3$ and NiAX$_3$ are found to be non-magnetic within DFT-D2, 
although they can stabilize magnetic phases unpon inclusion of a sufficiently large U at the metal atom sites.

The electronic structures predicted by density functional theory for these materials are sensitive to the choice 
of the Coulomb interaction model as manifested in the substantial differences in predicted DOS between 
the standard semi-local DFT-D2, and calculations including a local U correction.
Since the exchange interactions in the AFM phase are expected to vary inversely with the band gap
the approximations that underestimate the gap will overestimate the interaction strengths.
This sensitivity of the optimized ground-state results to the choice of the DFT approximation scheme which can  
alter the strength and range of the exchange interaction makes it desirable to benchmark the results against experiment in order to establish 
theoretical models upon which we can make further predictions of material's properties. 
We have found a variety of different stable and meta-stable magnetic configurations in single layer MAX$_3$ compounds.
The critical temperatures for the magnetic phases in the single layer limit are expected to be lower 
than in the bulk due to the reduction in the number of close neighbor exchange interactions.
We analyzed the magnetic phases of the 2D MAX$_3$ compounds by building a model Hamiltonian 
with exchange coupling parameters extracted by mapping the total energies from our {\em ab initio} calculations 
onto an effective classical spin model 
and obtained the critical temperatures through a statistical analysis based on the Metropolis algorithm~\cite{newman}.
The calculated critical temperatures assumed an Ising model that provides an upper bound for the expected
critical temperatures and were found to vary widely 
ranging between a few tens to a few hundred Kelvin.

Control of magnetic phases by varying the electric field in a field effect transistor device
is a particularly appealing strategy for 2D magnetic materials.
Our calculations indicate that a transition between AFM and FM phases
can be achieved by inducing carrier densities in 2D MAX$_3$ compounds
which are effectively injected through field effects using high $\kappa$-dielectrics, 
using ionic liquids, or by interfacing with ferroelectric materials. 
For materials exhibiting magnetic phases we find that use of the heavier chalcogen Te atoms
can reduce the carrier densities required for magnetic transitions.

The interdependence between atomic and electronic structure suggest that strains can be employed to tune magnetic phases, 
and they can be used to facilitate switching the magnetic configurations when they are combined together with carrier density variations.  
We have found that the ground state magnetic configuration can undergo phase transitions 
driven by in-plane compression or expansion of the lattice constants as small as a few percents in certain cases. 

Based on our calculations, we conclude that single layer MAX$_3$ transition metal trichalcogenides
are interesting candidate materials for 2D spintronics.
Their properties, including their magnetic transition temperatures, can be adjusted by the application
of external strains or by modifying the carrier densities in field effect transistor devices.
For monolayer CrSiTe$_3$ we showed in Fig.~\ref{carrier-str-dos-tc}(c) that the T$_c$ values can undergo an order of magnitude enhancement 
when subject to large carrier doping or in the presence of compressive or expansive strains. 
The sensitivity of these systems to variations in system parameters, such as composition,
details of the interface, and the exchange coupling of the magnetic properties with external fields,
offers ample room for future research that seek new functionalities in magnetic 2D materials.

\section{Acknowledgements}
We are thankful to the assistance from and computational resources provided 
by the Texas Advanced Computing Centre (TACC). 
We acknowledge financial support from NRF-2017R1D1A1B03035932 for BLC,  
the 2017 Research Fund of the University of Seoul for JJ, 
NRF-2014R1A2A2A01006776 for EHH, 
DOE BES Award SC0012670, and Welch Foundation 
grant TBF1473 for AHM.


\clearpage

\begin{center}
{\bf \large Supplemental Material}
\end{center}
In this supplement we present calculation results obtained within DFT-D2+U for the results discussed in the main text within DFT-D2. 
These comprise the total energy difference, lattice constants, the J-Coupling parameters the magnetic moments, transition temperatures, 
band structures, the associated density of states and the orbital projected density of states (PDOS) for selected compounds calculated for self-consistently for a different magnetic configurations.
We generally use the finite onsite repulsion U=4~eV and larger values for a few select compounds. 
The temperature dependence of the heat capacity is obtained through the Metropolis Monte Carlo simulation in a 32$\times$64 lattice.

\begin{figure}[htb!]
\begin{center}
\includegraphics[width=8cm]{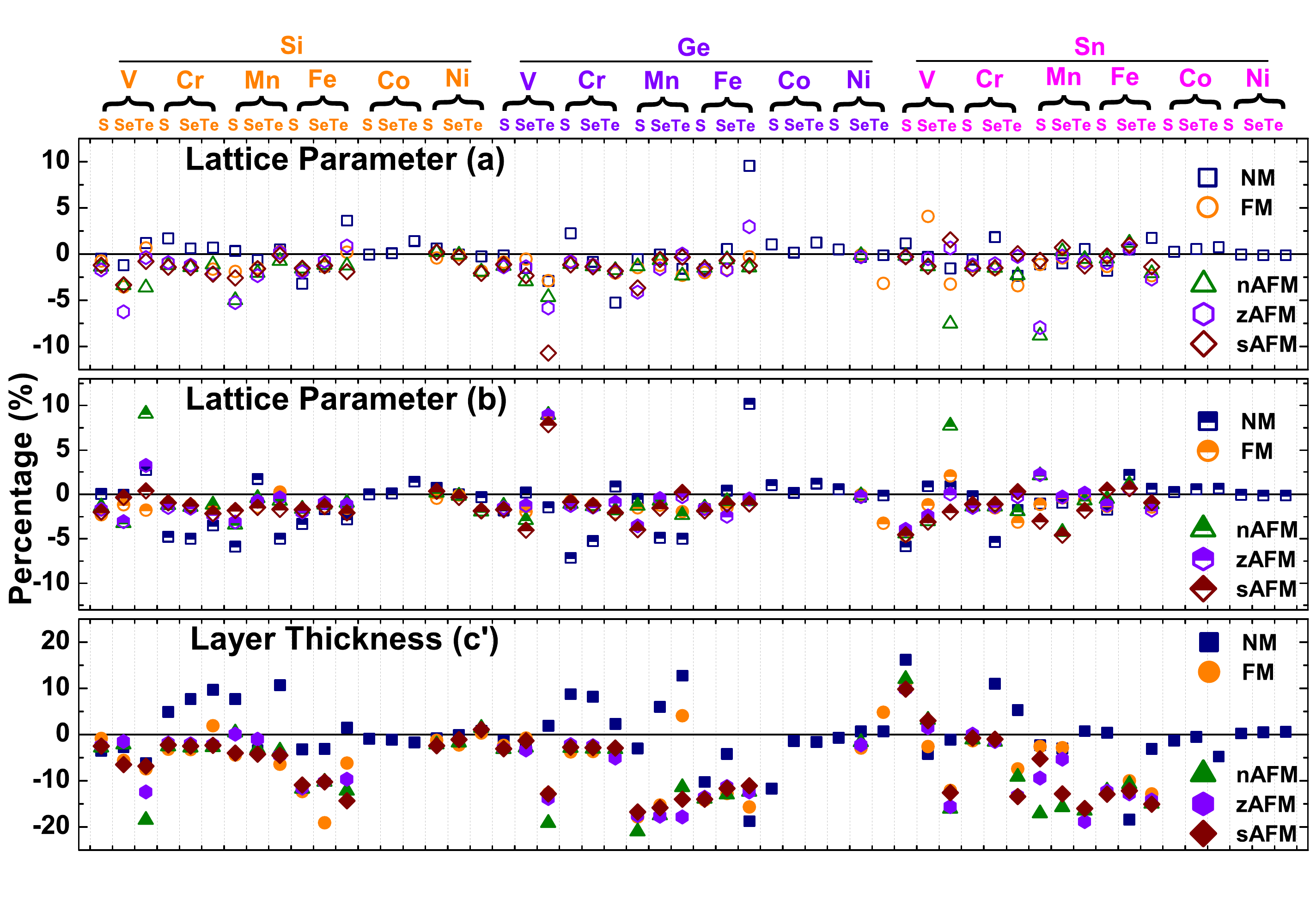}
\caption{(Color online) 
The percentage of change of the lattice parameters when relaxed within DFT-D2+U. In-plane parameters $a$ and $b$ define the rectangular unit cell and the layer thickness $c'$ is defined as the vertical distance between the chalcogen atoms from top to bottom layer in single layer MAX$_3$. The percentage of change is measured with reference to the structural parameters obtained within DFT-D2 to account for differences introduced by DFT-D2+U. The negative (positive) values indicates compression (expansion) of DFT-D2+U results with respect to DFT-D2.
}
\label{fig:lattice}     
\end{center}
\end{figure}

\begin{table*}[htb!]
\caption{Total energy relative to the lowest energy magnetic configuration
among ferromagnetic (FM), Neel antiferromagnetic (nAFM), Zigzagl antiferromagnetic (zAFM), Stripy antiferromagnetic (sAFM) and non-magnetic (NM) states in single layer transition metal trichalcogenide MSiX$_3$.  The absence of an entry for 
a magnetic configuration means that the corresponding state is not metastable.  Energies are  in eV/unit cell units within DFT-D2 and DFT-D2+U(=4eV). In the cases of CoASe$_3$ and CoATe$_3$ (A=Ge, Sn), the U parameter values are 6 and 5 eV, respectively. In the case of CoSnS$_3$ the U parameter is 7eV.
}
\begin{center}
\begin{tabular}{cccccc|ccccc}\\  \hline \hline
                     &   \multicolumn{5}{c|}{DFT-D2}  &  \multicolumn{5}{c}{DFT-D2+U}   \\ \hline
      M                &{NM }& {FM } &{nAFM }&{zAFM} &{sAFM } &{NM }& {FM } &{nAFM }&{zAFM} &{sAFM } \\ \hline
VSiS$_3$ &0	           &-1.1613	&-1.2972	&-1.3303	&-1.3437 &0	&-7.0070	&-7.1120	&-7.0640	&-7.0813 \\
VSiSe$_3$ &0	&-1.5713	&-1.4996	&-1.6140	&-1.5369 &0	&-6.3430	&-6.2205	&-6.1321	&-6.4343\\
VSiTe$_3$ &0	&-1.0129	&-0.7277	&-0.9473	&-0.8344 &0	&-7.6430	&-7.6448	&-7.6871	&-7.7340 \\ \hline
CrSiS$_3$ &0	&-3.8861	&-4.1986	&-4.0424	&-4.1240 &0	&-11.469	&-11.528	&-11.509	&-11.488 \\
CrSiSe$_3$&0	&-4.5213	&-4.6568	&-4.6038      &-4.6264 &0	&-12.346	&-12.305	&-12.377	&-12.284 \\
CrSiTe$_3$&0	&-4.9580	&-4.8682	&-4.9644	&-4.9308 &0	&-12.818	&-12.707	&-12.847	&-12.677 \\ \hline
MnSiS$_3$ &0	&-2.8011	&-2.8346	&-2.7147	&-2.8214 &0	&-11.638	&-11.534	&-11.500	&-11.533 \\
MnSiSe$_3$ &0	&-3.5876	&-3.4767	&-3.3192	&-3.5827 &0	&-13.941	&-13.579	&-13.462	&-13.828 \\
MnSiTe$_3$ &0	&-3.6131	&-3.5226	&-3.4269	&-3.6211 &0	&-12.772	&-12.398	&-12.386	&-12.676 \\ \hline
FeSiS$_3$ &0	&-1.8437	&-1.8710	&-1.8764	&-1.8939 &0	&-3.5815	&-3.8281      &-3.9170	&-3.6311 \\
FeSiSe$_3$ &0	&-2.7585	&-2.8108	&-2.8106	&-2.8068 &0	&-5.4672	&-5.7992	&-5.8908	&-5.5514 \\
FeSiTe$_3$ &0	&-2.0942	&-2.3038	&-2.1277	&-2.3008 &0	&-5.5681	&-5.9098	&-5.9631	&-5.5911 \\ \hline
CoSiS$_3$ &0	&-	          &-	          &-	          &-             &0	&1.2045	&1.3376	&1.5138	&1.2844\\
CoSiSe$_3$ &0	&-	          &-	          &-	          &-             &0	&-0.6407	&-0.4611	&-0.5951	&-0.5135\\
CoSiTe$_3$ &0	&-	          &-	          &-	          &-             &0	&-0.1856	&-0.1892	&-0.2751	&-0.2324\\ \hline
NiSiS$_3$ &0   	&-0.0603	&-0.0954	&-       	&-0.1367 &0	&-0.9432	&-0.8072	&-0.6022	&-0.8240 \\
NiSiSe$_3$ &0	&-0.0720	&-0.0569	&-	          &-0.0470 &0	&-0.6965	&-0.5416	&-0.3322	&-0.5701 \\
NiSiTe$_3$ &0	&-1.3563	&-1.3465	&-                 &-1.3455 &0	&-0.2383	&-0.1520	&-0.0791	&-0.1773 \\ \hline \hline
VGeS$_3$ &0	&-0.8551	&-1.0047	&-1.0497	&-1.0135 &0	&-6.8186	&-6.8887	&-6.8697	&-6.8730 \\
VGeSe$_3$ &0	&-1.2948	&-1.3801	&-1.4274	&-1.3887 &0	&-7.5519	&-6.3725	&-7.5883	&-6.3054 \\
VGeTe$_3$ &0	&-1.4374	&-1.3504	&-1.4537	&-1.4363 &0	&-7.6375	&-7.7589	&-7.7523	&-7.5149 \\  \hline
CrGeS$_3$ &0	&-3.5150	&-3.6804	&-3.6183	&-3.6490 &0	&-11.869	&-11.852	&-11.879	&-11.859 \\
CrGeSe$_3$ &0	&-4.3062	&-4.3465	&-4.3600	&-4.3538 &0	&-15.271	&-15.165	&-15.273	&-15.205 \\
CrGeTe$_3$ &0	&-4.5408	&-4.3951	&-4.4796	&-4.4643 &0	&-12.411	&-12.220      &-12.422	&-12.283 \\  \hline
MnGeS$_3$ &0	&-2.0702	&-2.1643	&-2.0776	&-2.1587 &0	&-11.569	&-11.373	&-11.328	&-11.324 \\
MnGeSe$_3$ &0	&-3.3436	&-3.2496	&-3.2720	&-3.2851 &0	&-12.204	&-11.978	&-11.841	&-12.023 \\
MnGeTe$_3$ &0	&-2.5526	&-2.4758	&-2.4865	&-2.4937 &0	&-12.387	&-12.285	&-12.131	&-12.303 \\  \hline
FeGeS$_3$ &0	&-1.4965	&-1.5209	&-1.4891	&-1.4687 &0	&-2.9247	&-3.1896	&-3.3015	&-3.1439 \\
FeGeSe$_3$ &0	&-2.0193	&-2.0636	&-2.0460	&-2.0312 &0	&-4.8523	&-5.0595	&-5.1374	&-4.9942 \\
FeGeTe$_3$ &0	&-0.3953	&-0.5724	&-0.0668	&-0.5026  &0&-5.4433	&-5.6417	&-5.7546	&-5.6346 \\  \hline
CoGeS$_3$ &0	&-	          &-	          &-	          &-              &0&1.5376	&1.5652	&1.8399	&1.6939 \\
CoGeSe$_3$ &0	&-	          &-	          &-	          &-              &0&-0.1897	&-0.3510	&-0.513	&-0.2316	 \\
CoGeTe$_3$ &0	&-	          &-	          &-	          &-              &0&-0.1367	&-0.1056	&-0.1937	&-0.0757\\  \hline
NiGeS$_3$ &0	&-	          &-	          &-	          &-              &0&-0.8830	&-0.8023	&-0.5681	&-0.6580 \\
NiGeSe$_3$ &0	&-0.0080      &-0.0820      &-0.0065      &-              &0&-0.5482	&-0.4961	&-0.2754	&-0.3397 \\
NiGeTe$_3$ &0	&-0.0227      &-	          &-	          &-              &0&-0.7100	&-0.7052	&-	           &-	     \\  \hline  \hline
VSnS$_3$ &0	&-0.4671	&-0.3655	&-0.5517	&-0.5334 &0	&-7.4887	&-7.5444	&-7.5591	&-7.5447 \\
VSnSe$_3$ &0	&-0.6769	&-0.5615	&-0.7327	&-0.6769 &0	&-5.7249	&-6.8284	&-6.8784	&-6.8273 \\
VSnTe$_3$ &0	&-0.6710	&-0.2878	&-0.6709	&-0.4702 &0	&-6.2785	&-6.8289	&-6.8647	&-6.8498\\ \hline
CrSnS$_3$ &0	&-3.2215	&-3.1614	&-3.3685	&-3.2414 &0	&-10.146	&-9.9718	&-10.129	&-10.037 \\
CrSnSe$_3$ &0	&-3.5779	&-3.4652	&-3.7060	&-3.5659 &0	&-12.518	&-12.304	&-12.492	&-12.386 \\
CrSnTe$_3$ &0	&-2.8574	&-2.5545	&-2.7088	&-3.1483 &0	&-12.138	&-11.884	&-12.160	&-12.069 \\ \hline
MnSnS$_3$ &0	&-1.4988	&-2.0207	&-1.5716	&-1.8249 &0	&-10.538	&-10.764	&-10.751	&-10.396 \\
MnSnSe$_3$ &0	&-2.9251	&-2.9376	&-2.8605	&-2.9514 &0	&-11.725	&-11.776	&-11.579	&-11.723 \\
MnSnTe$_3$ &0	&-3.3821	&-3.3276	&-3.3181	&-3.3581 &0	&-10.060	&-9.9105	&-9.7974	&-9.7948 \\ \hline
FeSnS$_3$ &0	&-1.5815	&-1.3056	&-1.3138	&-1.5535 &0	&-3.5171	&-3.9264	&-3.8132	&-3.7811 \\
FeSnSe$_3$ &0	&-1.1891	&-1.3013	&-1.2461	&-1.2751 &0	&-6.4132	&-6.4131	&-6.2240	&-6.2525 \\
FeSnTe$_3$ &0	&-1.2698	&-1.2598	&-1.1353	&-1.1218 &0	&-5.6362	&-5.7804	&-5.7549	&-5.8944 \\ \hline
CoSnS$_3$ &0	&-	          &-	          &-	          &-             &0 &-0.1655	&-0.0500	&-0.1504	&0.4196 \\
CoSnSe$_3$ &0	&-	          &-	          &-	          &-             &0	&-0.2182	&-0.0590	&-0.2241	&-0.1952 \\
CoSnTe$_3$ &0	&-	          &-	          &-	          &-             &0	&-0.2923	&-0.1863	&-0.3729	&-0.3337 \\ \hline
NiSnS$_3$ &0	&-	          &-	          &-	          &-             &0 &-0.3865	&-0.4301	&-0.3665	&-0.3113 \\
NiSnSe$_3$ &0	&-	          &-	          &-	          &-             &0	&-0.2796	&-0.1472	&-0.2510	&-0.2328 \\
NiSnTe$_3$ &0	&-	          &-	          &-	          &-             &0	&-0.2747      &-	          &-                  &- \\ \hline  \hline
\end{tabular} \label{tab:sitotalenergies}
\end{center}
\end{table*}

\begin{figure*}[htbp!]
\begin{center}
\includegraphics[width=18cm]{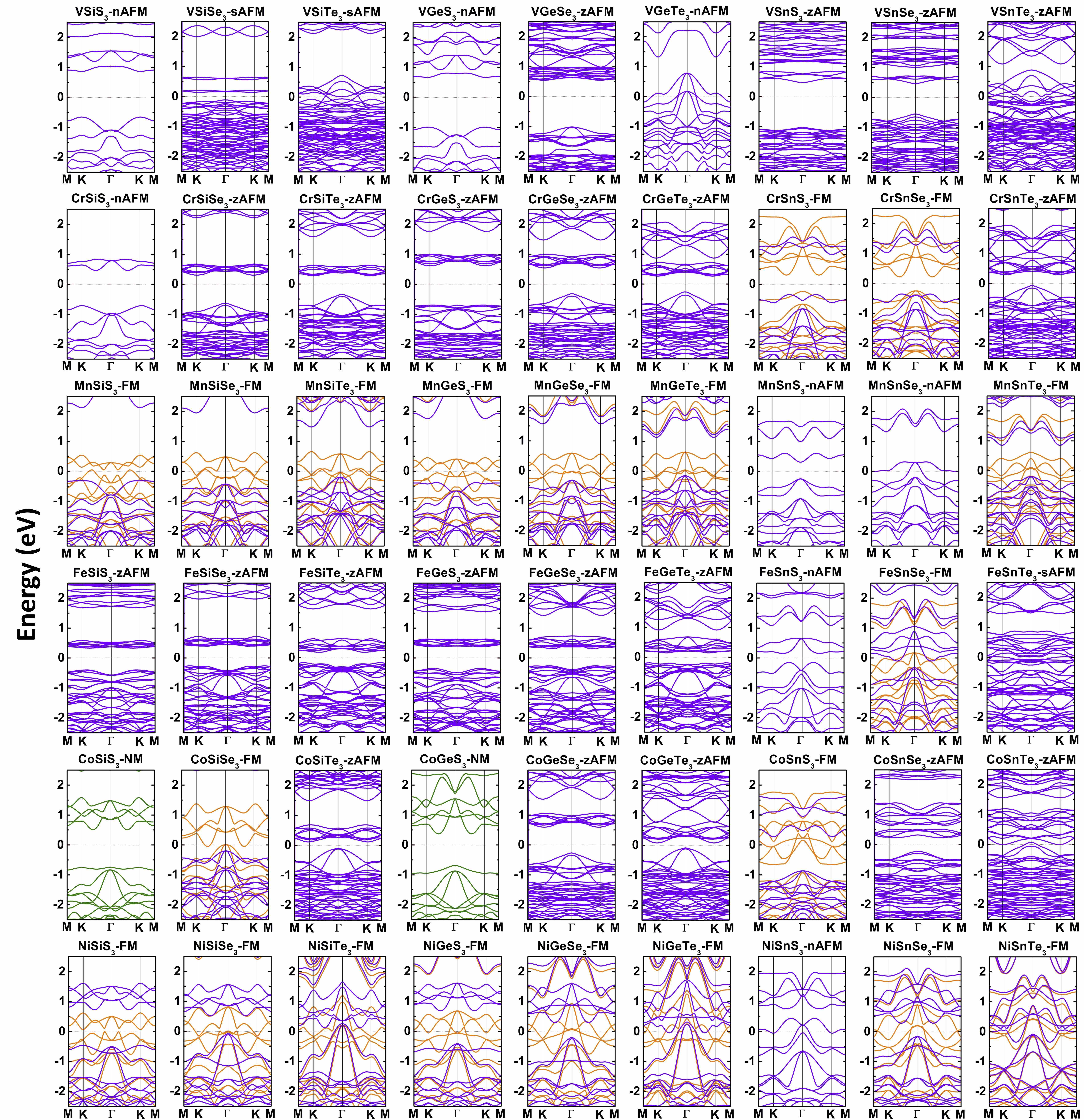}
\caption{(Color online) 
The DFT-D2+U band structures for single-layer MAX$_{3}$ compounds in their lowest-energy magnetic configurations for M = V, Cr, Mn, Fe, Co and Ni transition-metal atoms with combination of A =  Si, Ge, Sn and X = S, Se, Te chalcogen atoms. The plotted band structures were calculated using the triangular structural unit cell, except for the cases of sAFM and zAFM, which are predicted to have a larger periodicity magnetic structure
for which we used a triangular unit cell with doubled lattice constant. The bands are violet for AFM configurations, violet and orange for the down and up split spin bands in the FM configurations, and green for the NM phases.}
\label{BS}
\end{center}
\end{figure*}

\begin{sidewaysfigure*}[htbp!]
\centering
\textcolor{white}{\rule{6.4in}{3.6in}}
\includegraphics[width=24cm]{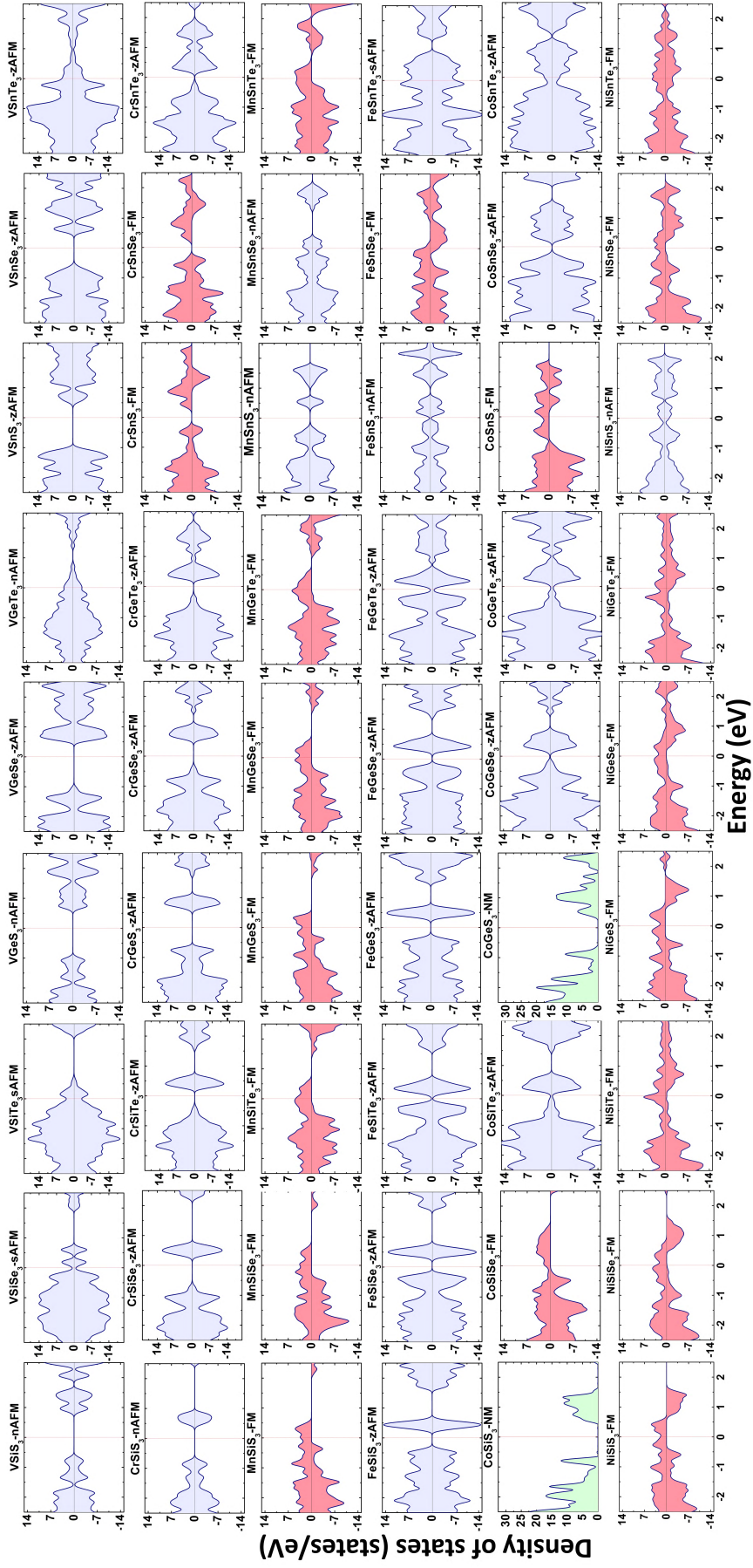}
\caption{(Color online) 
The DFT-D2+U density of states (DOS) for single-layer MAX$_{3}$ compounds in their lowest-energy magnetic configurations for M = V, Cr, Mn, Fe, Co and Ni transition-metal atoms with combination of A =  Si, Ge, Sn and X = S, Se, Te chalcogen atoms. The plotted DOS were calculated using the triangular structural unit cell, except for the cases of sAFM and zAFM, which are predicted to have a larger periodicity magnetic structure that has triangular unit cell with a doubled lattice constant. The state are grey for AFM configurations, red for states in the FM configurations, and green for the NM phases.}
\label{tri-DOS}
\end{sidewaysfigure*}

\begin{figure*}[htb!]
\begin{center}
\includegraphics[width=18cm]{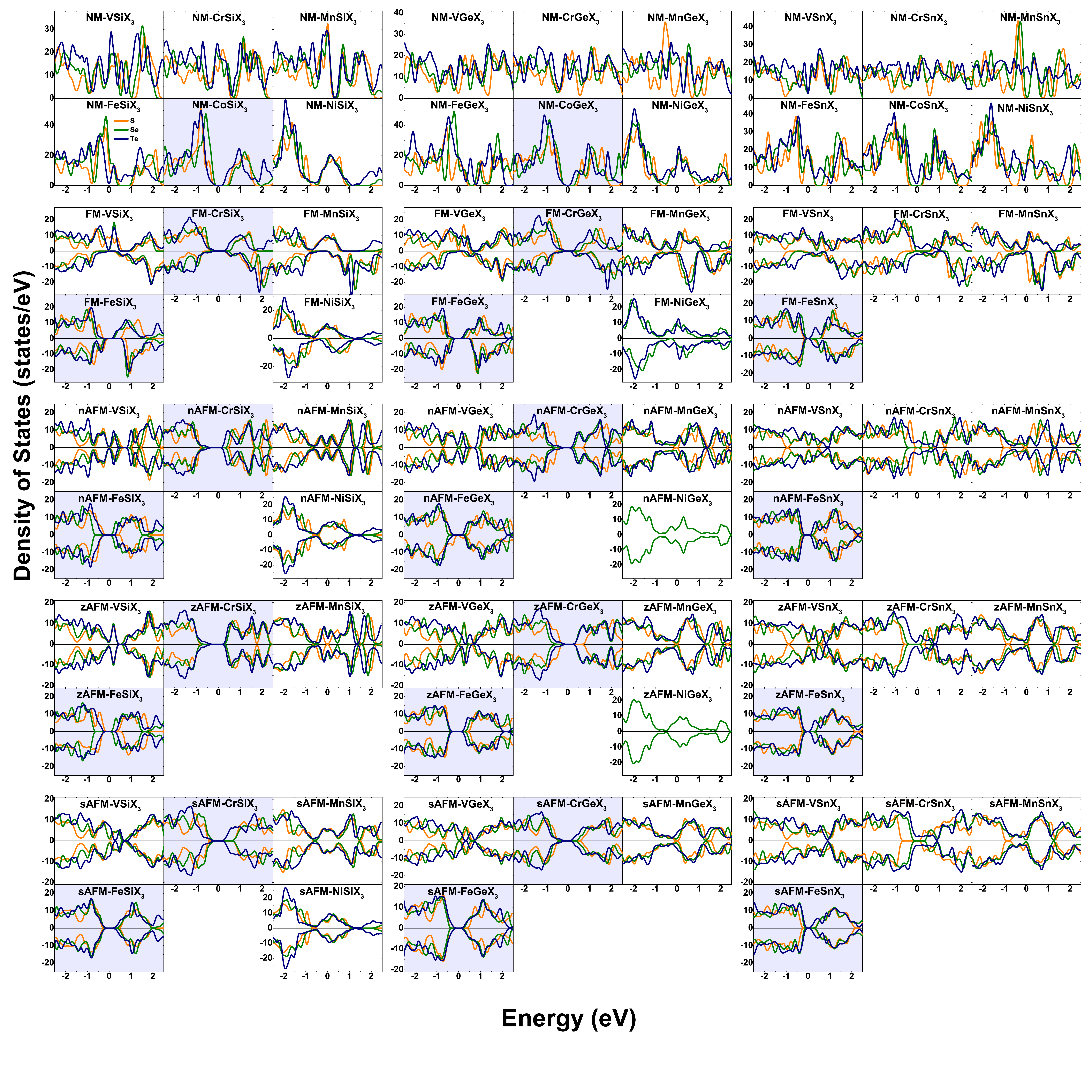}
\caption{(Color online) Total density of states (DOS) for the magnetic and non-magnetic ground-states of MAX$_3$ within DFT-D2 obtained using a rectangular unit cell. The Fermi energy is positioned at $E=0$.} 
\label{fig:dos}
\end{center}
\end{figure*}

\begin{figure*}[hp!]
\begin{center}
\includegraphics[width=18cm]{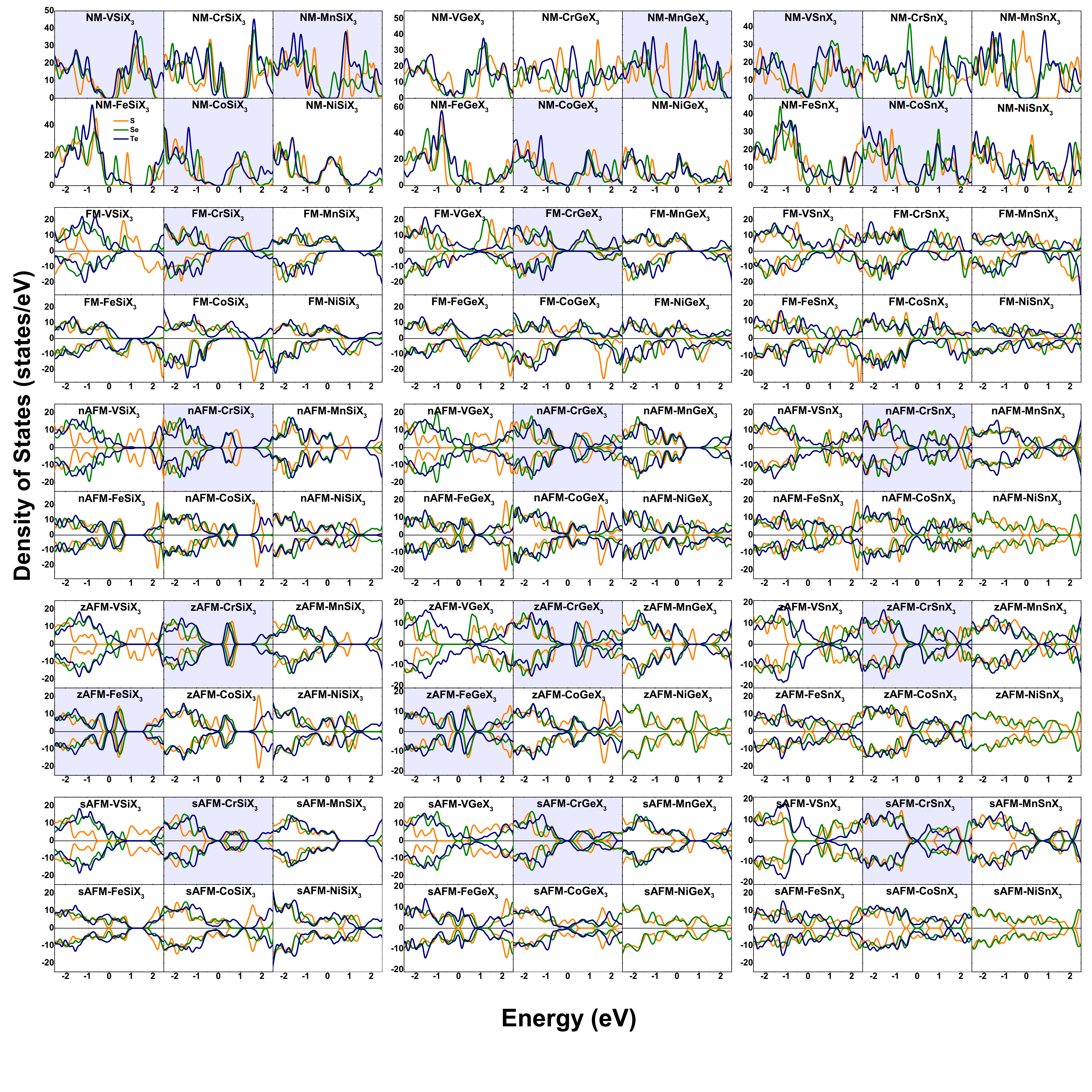}
\caption{(Color online) Total density of states (DOS) for the magnetic and non-magnetic ground-states of MAX$_3$ within DFT-D2+U using rectangular lattice. In the cases of CoASe$_3$ and CoATe$_3$ (A=Ge, Sn), the U parameter values are 6 and 5 eV, respectively. In the case of CoSnS$_3$ the U parameters is 7eV.
We have placed the Fermi energy at $E=0$.} 
\label{fig:dos}
\end{center}
\end{figure*}

\begin{figure*}[htbp!]
\begin{center}
\includegraphics[width=16cm]{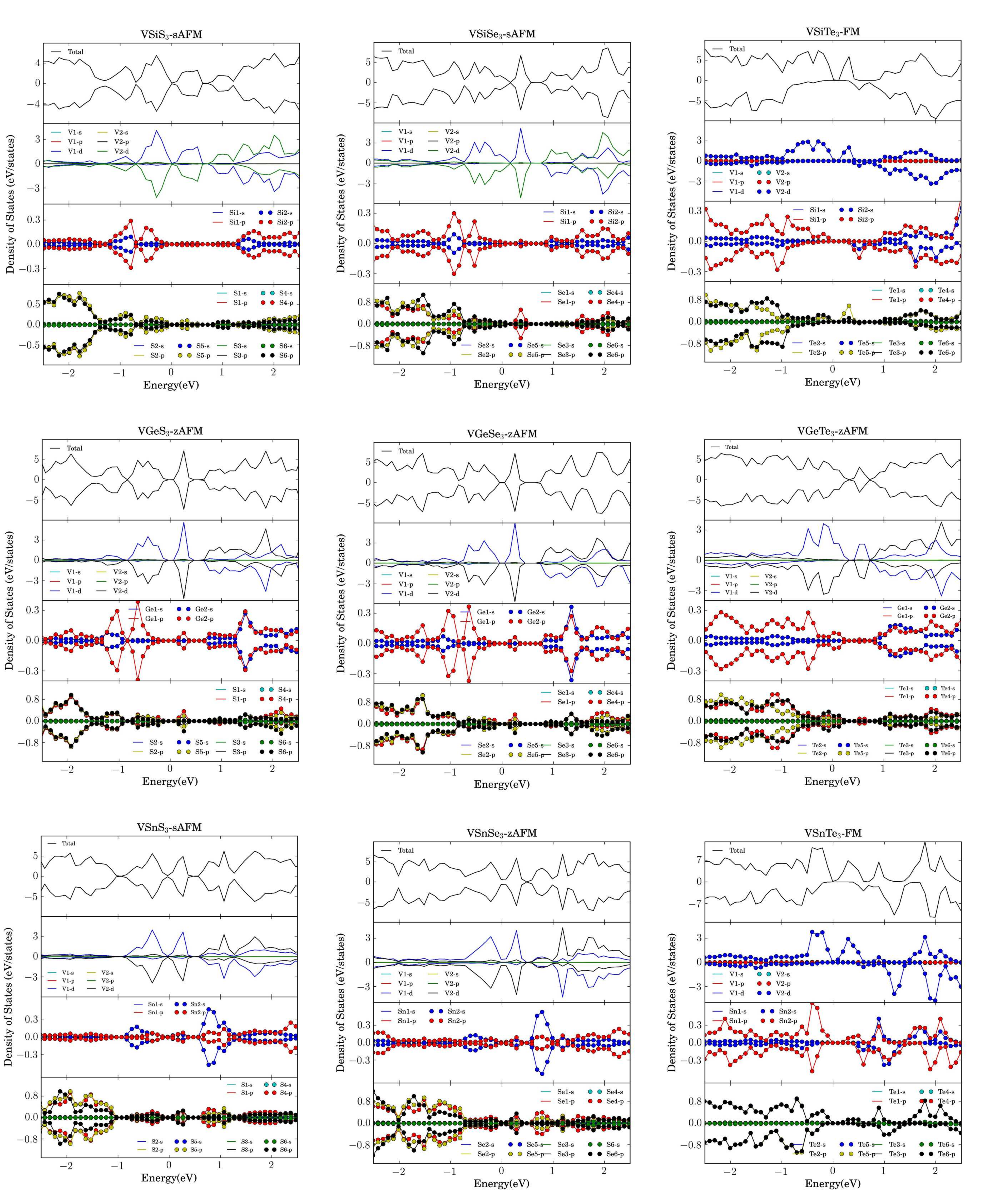}
\caption{(Color online) Total and projected density of states (PDOS) for the ground state of VAX$_3$ obtained within DFT-D2. The Fermi energy is positioned at $E=0$.} 
\label{fig:pdos-vax3}
\end{center}
\end{figure*}

\begin{figure*}[htbp!]
\begin{center}
\includegraphics[width=16cm]{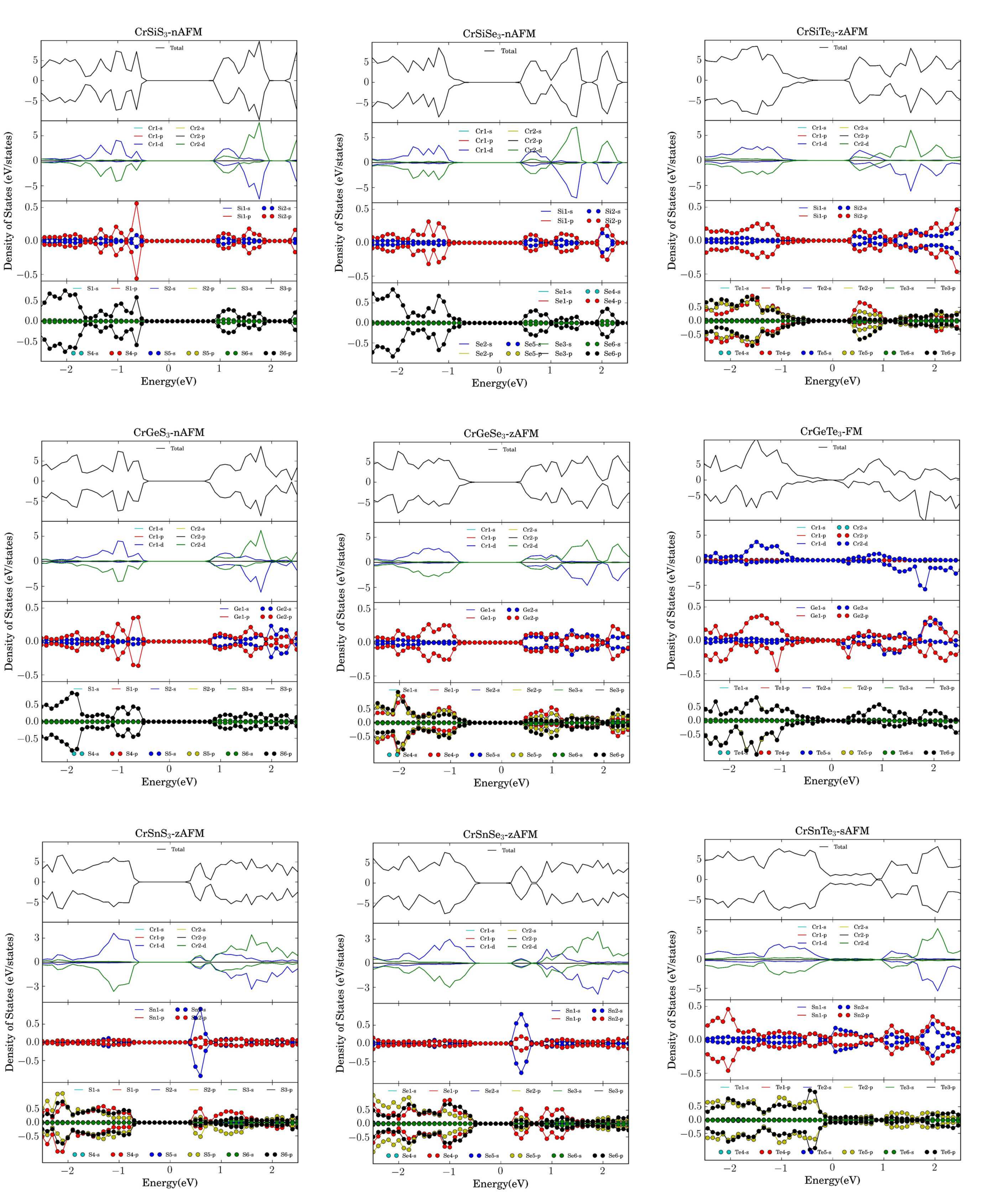}
\caption{(Color online) Total and projected density of states (PDOS) for the ground state of CrAX$_3$ obtained within DFT-D2. The Fermi energy is positioned at $E=0$.} 
\label{fig:pdos-crax3}
\end{center}
\end{figure*}

\begin{figure*}[htbp!]
\begin{center}
\includegraphics[width=16cm]{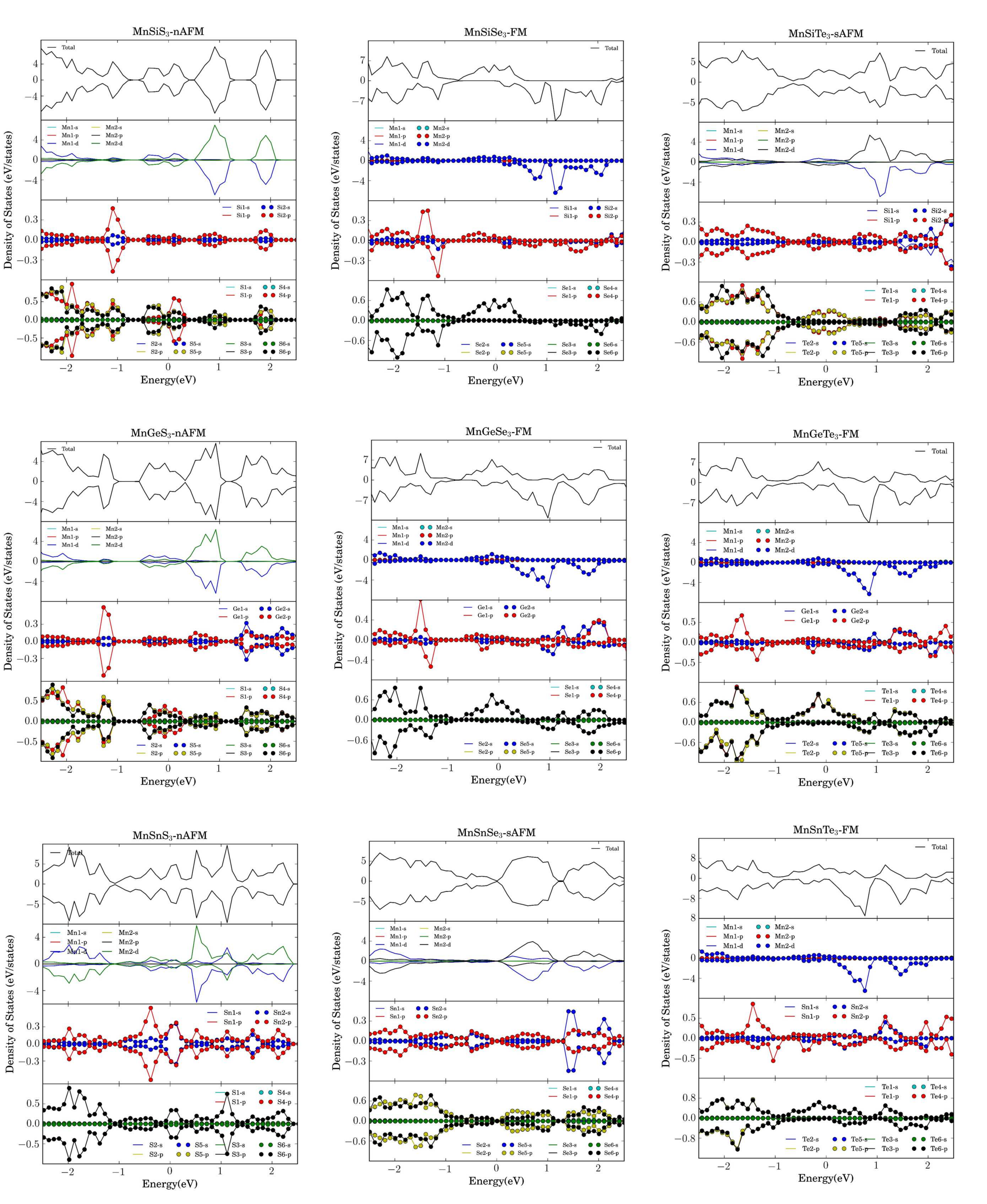}
\caption{(Color online) Total and projected density of states (PDOS) for the ground state of MnAX$_3$ obtained within DFT-D2. The Fermi energy is positioned at $E=0$.} 
\label{fig:pdos-mnax3}
\end{center}
\end{figure*}           

\begin{figure*}[htbp!]
\begin{center}
\includegraphics[width=16cm]{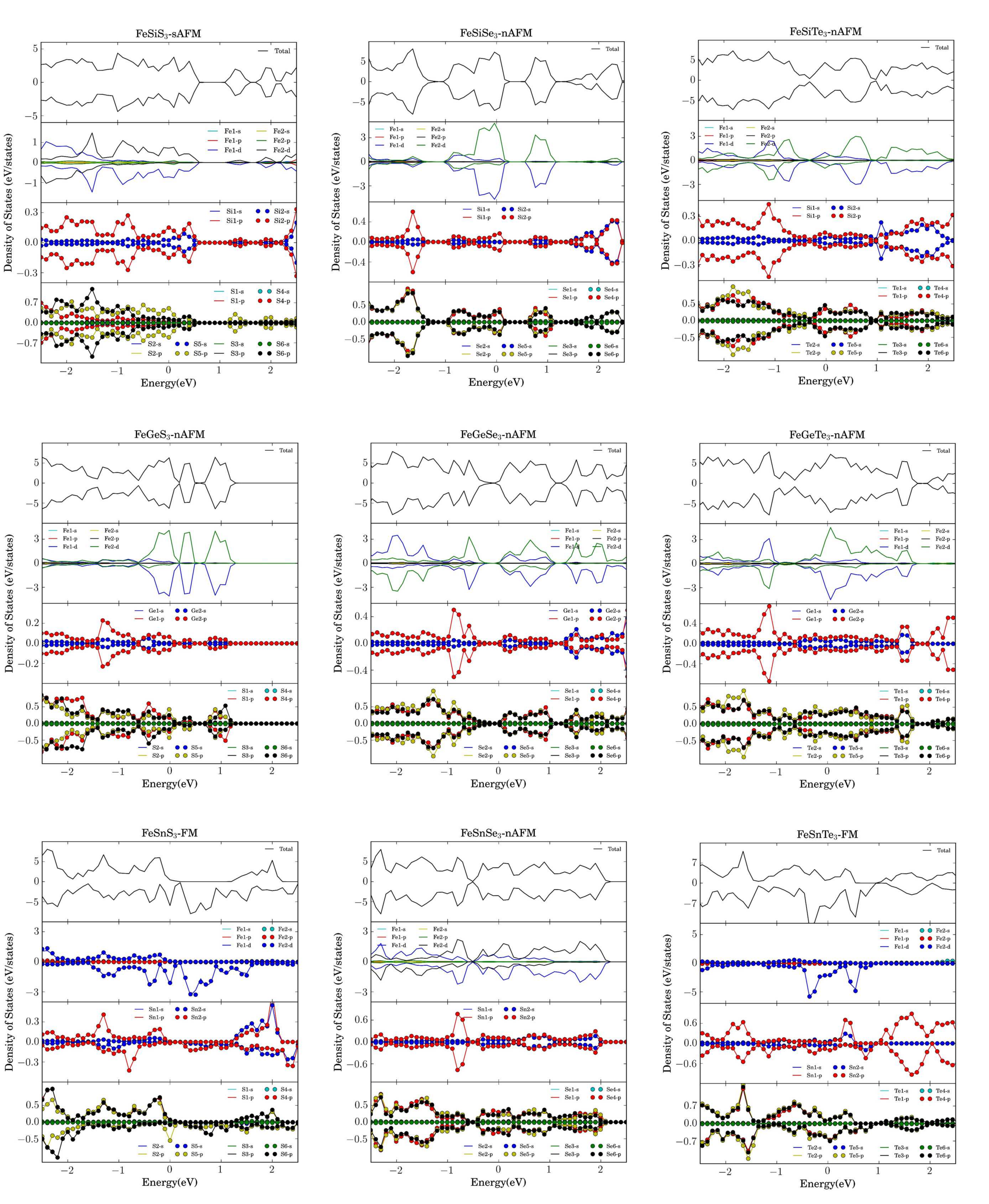}
\caption{(Color online) Total and projected density of states (PDOS) for the ground state of FeAX$_3$ obtained within DFT-D2. The Fermi energy is positioned at $E=0$.} 
\label{fig:pdos-feax3}
\end{center}
\end{figure*}

\begin{figure*}[htbp!]
\begin{center}
\includegraphics[width=16cm]{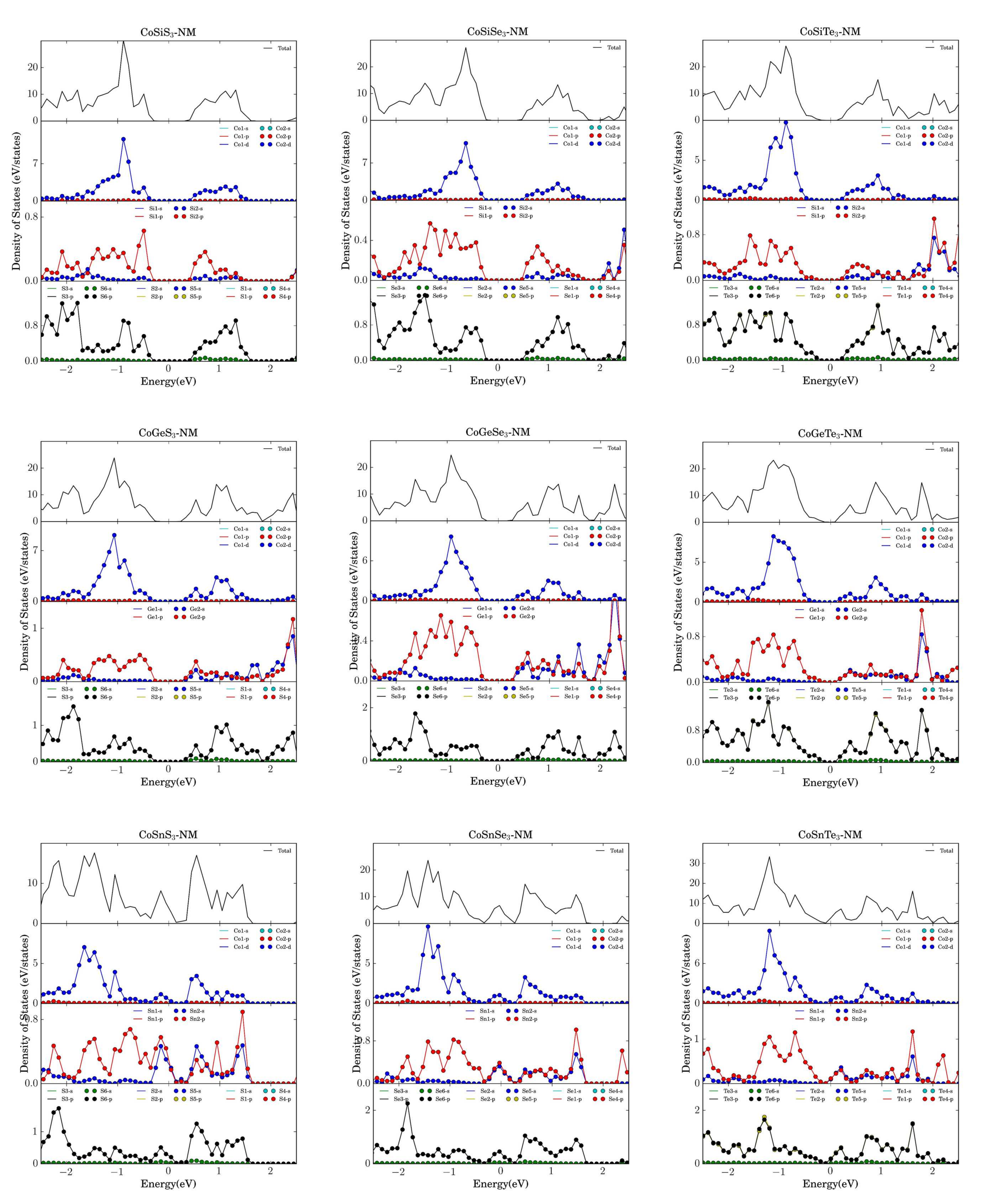}
\caption{(Color online) Total and projected density of states (PDOS) for the ground state of CoAX$_3$ obtained within DFT-D2. The Fermi energy is positioned at $E=0$.} 
\label{fig:pdos-coax3}
\end{center}
\end{figure*}

\begin{figure*}[htbp!]
\begin{center}
\includegraphics[width=16cm]{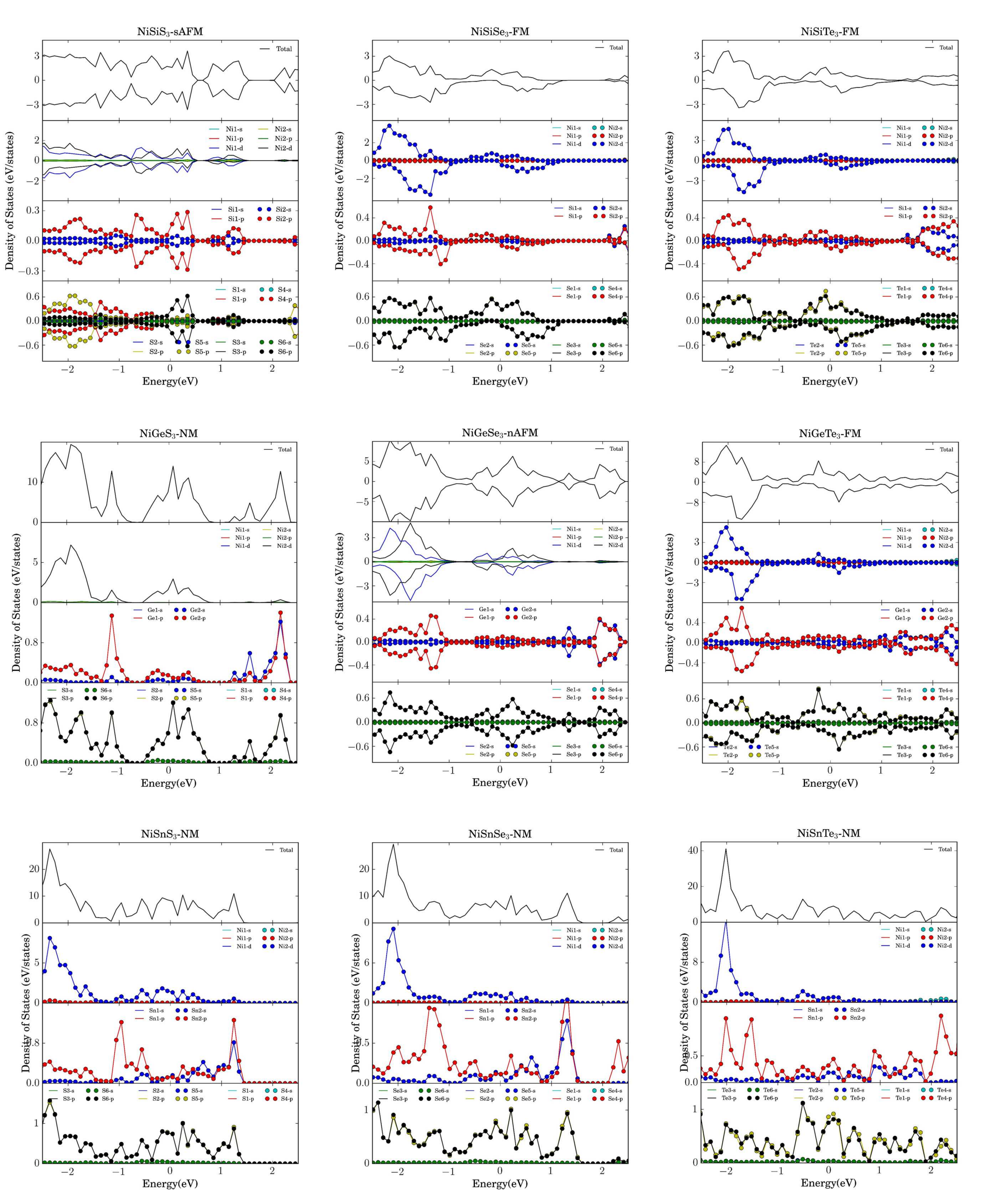}
\caption{(Color online) Total and projected density of states (PDOS) for the ground state of NiAX$_3$ obtained within DFT-D2. The Fermi energy is positioned at $E=0$.} 
\label{fig:pdos-niax3}
\end{center}
\end{figure*}

\begin{figure*}[htbp!]
\begin{center}
\includegraphics[width=16cm]{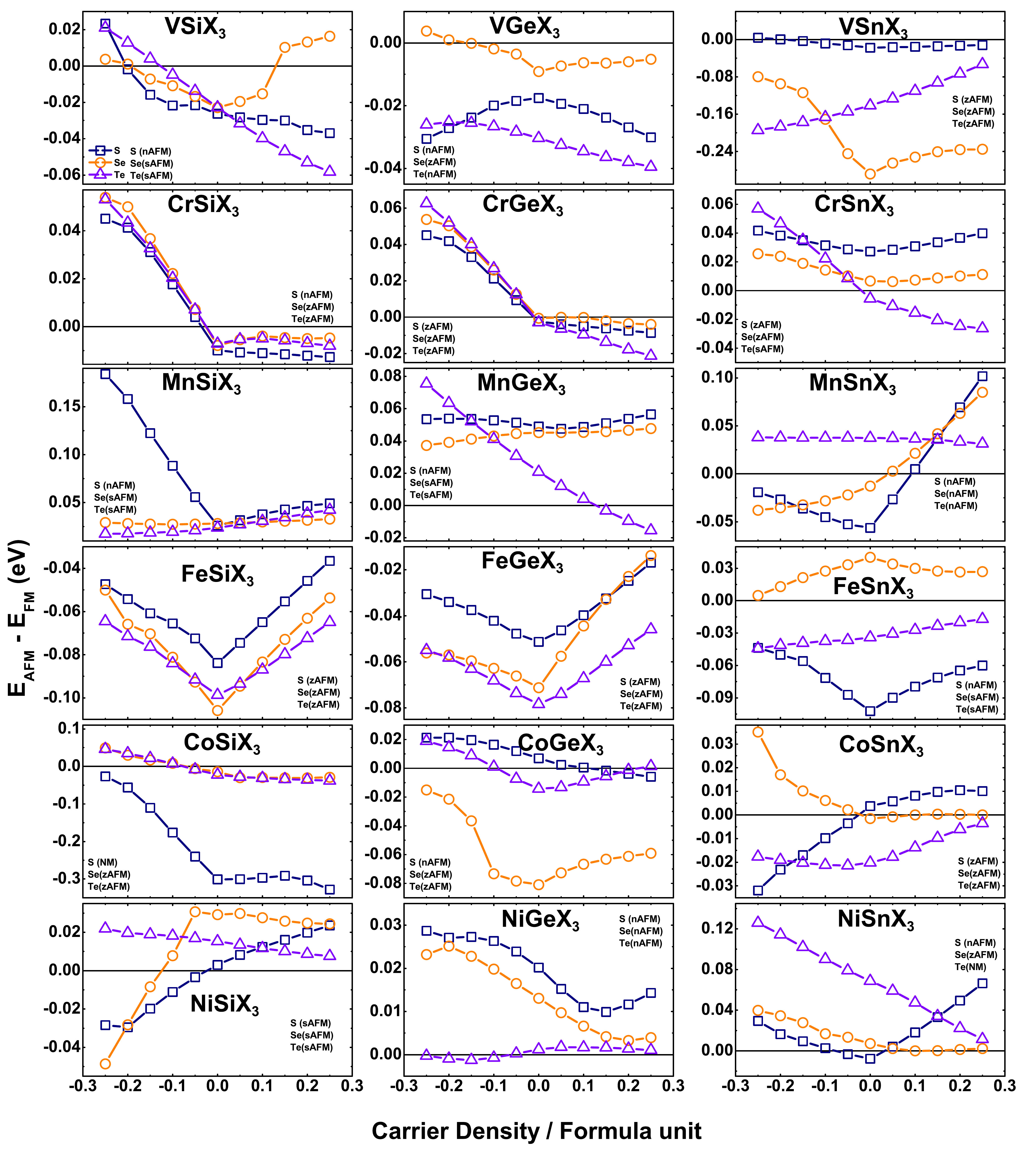}
\caption{(Color online) 
Carrier density dependent total energy differences per MAX$_3$ formula unit 
between the AFM and FM states of V, Cr, Mn, Fe,Co, Ni based single layer trichalcogenides calculated within DFT-D2+U.  
The AFM ground-states favored near charge neutrality can often be switched to FM states at accessible
carrier densities in V, Mn, Ni, and Fe based compounds. 
Densities of up to a few $ { \sim10^{14}}$ electrons per ${\rm cm}^2$ should be in principle accessible through ionic liquid or gel gating.
A carrier density of 0.1 electrons per MAX$_3$ formula unit corresponds to $\sim$$ {6\times 10^{13}}$ electrons per cm$^2$ when the lattice constant is $\sim$6 angstroms.
}
\label{carrier_magnetism_U}
\end{center}
\end{figure*}

\begin{table*}[htb!]
\caption{The theoretical two dimensional lattice parameters for 
transition metal trichalcogenide MAX$_3$ compounds.
These results were obtained using DFT-D2 and are expressed in $\AA$; 
a is the in-plane lattice constant and c$'$ is the layer thickness,  
{\it i..e} it is the distance between the planes
containing the three chalcogen atoms in a single MAX$_3$ layer. }
\begin{center}
\begin{tabular}{cc|ccc|ccc|ccc|ccc|ccc}\\ \hline \hline
 MAX$_3$   &  & &{NM}&&&{ FM }&&& {nAFM}&&&{ zAFM }&&& {sAFM}& \\ \hline
                   &  &a(\AA) &b(\AA)& c$'$(\AA) &a(\AA)&b(\AA) & c$'$(\AA) &a(\AA) &b(\AA)   &c$'$(\AA)  &a(\AA)&b(\AA)  & c$'$(\AA) &a(\AA)&b(\AA) & c$'$(\AA) \\ \hline
VSiS$_3$    &   &5.992 &9.899 &3.284   &5.997 &10.228&3.154  &5.952 &10.306    &3.097   &5.940 &10.272  &3.118   &5.960 &10.252 & 3.108 \\ 
VSiSe$_3$    &  &6.267  &10.451 &3.580    &6.272 &10.733 &3.415  &6.286 &10.894 &3.289 &6.159  &10.874    &3.406  &6.303  &10.821  &3.317   \\ 
VSiTe$_3$    &   &6.972 &11.657 &3.796    &6.822 &11.657 &3.651 &6.933&12.011 &3.452 &6.734&11.983  &3.504 &6.778 &11.621  &3.689    \\  \hline
CrSiS$_3$    &  &5.968&9.697 &3.201  &5.902 &10.223  &3.097 & 5.858&10.147 & 3.136    &5.914&10.118  & 3.181 &5.841&10.219  &3.149   \\ 
CrSiSe$_3$    &  &6.242&10.236 &3.519   &6.232 & 10.794 &3.281 &6.189 &10.720 &3.321    &6.240 &10.694  &3.368  &6.178 &10.787  & 3.333    \\ 
CrSiTe$_3$    &  &6.790 &11.277 &3.799   &6.843 &11.853  &3.631 &6.829 & 11.823&3.520   &6.826 &11.719  &3.541  &6.823 &11.728  & 3.539  \\  \hline
MnSiS$_3$    &  &5.910 &9.639 & 3.149  &5.951 &10.322  &3.148 &5.904 & 10.381& 3.277  &5.897 &10.412  &3.298  &5.910 &10.331  &3.173    \\ 
MnSiSe$_3$    &  &6.237 &10.114 &3.449   &6.270 &10.876  &3.355 &6.216 &10.944 & 3.410    &6.214 &10.931  &3.561  &6.233 &10.884  &3.378    \\ 
MnSiTe$_3$    &  &6.837 &11.224 &3.611  &6.922 &11.985  &3.497 &6.967 & 11.582    &3.648  &6.813 &11.895  &3.721  &6.869 &11.890 &3.571    \\  \hline
FeSiS$_3$     &  &5.852 &10.136 &2.744  &5.901  &10.224  &2.911 &5.920 &10.251 &2.914    &5.904 & 10.242 & 2.916 &5.918 &10.239  &2.937    \\ 
FeSiSe$_3$    &  &6.292 &10.889 &2.920  &6.267  &10.851  &2.920 &6.288 &10.885 &3.123    &6.289 &10.884  &3.124  &6.272 &10.866  & 3.130    \\ 
FeSiTe$_3$    &  &6.766 &10.976 &3.536   &6.947 &11.816  &3.536 &6.846 &11.888 & 3.278   &6.951 &11.907  &3.320  &6.803 &11.754  &3.279   \\  \hline
CoSiS$_3$    &  &5.735 &9.934 &2.874  &-&-&-&- &-  &-&- &-    &-  &-  &-  &-    \\ 
CoSiSe$_3$    &  &6.076 &10.524 &3.031   &-&-&-&- &-  &-&- &-    &-  &-  &-  &-    \\ 
CoSiTe$_3$    & &6.702 &11.612 &3.134   &-&-&-&- &-  &-&- &-    &-  &-  &-  &-     \\  \hline
NiSiS$_3$    &  & 5.754 &9.960 &2.976  &5.695  &9.871  &3.040 &5.752 &9.965 & 2.989  & - &-  &-  &5.751  &9.970  &2.993    \\ 
NiSiSe$_3$    &  &6.027  &10.432 &3.161   &6.027  &10.432  &3.170 &6.027  &10.432 &  3.170  & -& - &-  &6.013 &10.420  &3.197    \\ 
NiSiTe$_3$    & &6.558  &11.341 &3.319  &6.441  &11.150  &3.393 &6.441 &11.150 &3.393  & -  &-  & - &6.441 &  11.150 & 3.393   \\  \hline \hline
VGeS$_3$    &  &6.153 &9.987 &3.260  &6.080  &10.465  &3.111 &6.045  &10.470 &3.093    &6.050 &10.443  &3.104  &6.056 &10.438  &3.098    \\ 
VGeSe$_3$    &  &6.328 &10.580 &3.613   &6.425 &10.966  &3.354 &6.373 &11.044 &3.286  &6.371   &11.025  & 3.286 &6.380 &11.000  & 3.294  \\ 
VGeTe$_3$    &  &6.743 &11.444 &3.964   &7.022  &11.937  & 3.596 &6.904 &11.967 &3.496    &6.835 &11.963  &3.671  &6.809 &11.709  &3.640   \\  \hline
CrGeS$_3$    &  &6.106 &9.693 &3.284  &5.999 &10.392  &3.083 &5.952 &10.309 &3.133    &6.000 & 10.285 &3.174 &5.941  &10.373  &3.144     \\ 
CrGeSe$_3$    &  &6.260 &10.388 &3.569   &6.311 &10.932  &3.277 &6.266 &10.854  &3.326  &6.310 &10.835 &3.366  &6.259 &10.909  &3.336    \\ 
CrGeTe$_3$    &  &6.679 &11.254 &3.908  &6.887  &11.929  &3.451&6.862 &11.886 &3.511  & 6.893  &11.940  & 3.461 &6.852 &11.914  &3.511     \\  \hline
MnGeS$_3$    &  &6.000 &10.394 &2.517   &6.043 &10.466  &2.812 &6.051 &10.480  &2.744   & 6.030 &10.490   &2.811  &6.048 &10.459  &2.816     \\ 
MnGeSe$_3$    & &6.325  &10.451 &3.252   &6.354 &11.006  &3.046 &6.360 &11.017 &3.013   &6.334  &11.045  &3.048  &6.361 &10.994  &3.054     \\ 
MnGeTe$_3$    &  &6.802 &11.389 &3.655   &6.881 &11.943  &3.327 &6.843 &11.877 &3.354    &6.861 &12.029  &3.296  &6.900 &11.949  & 3.335    \\  \hline
FeGeS$_3$     &  &5.977 &10.352 &2.529   &5.996 &10.391  &2.873 &6.014 &10.421 & 2.864  &5.997  &10.408  &2.880 &6.011  &10.394  &2.865     \\ 
FeGeSe$_3$    &  &6.439 &11.143 &2.853   &6.329 & 10.957 &3.075 & 6.388& 11.048& 3.083   &6.329 &10.985  &3.088  &6.375 &11.012  &3.098    \\ 
FeGeTe$_3$     &  &7.282 &12.785 &2.970   &6.913 &11.997  &3.256 &6.881 & 11.981& 3.250 & 7.127  &12.045  &3.282 &6.910  &11.999  & 3.281   \\  \hline
CoGeS$_3$    &  &5.887 &10.197 &2.578   &-&-&-&- &-  &-&- &-    &-  &-  &-  &-    \\ 
CoGeSe$_3$      &  &6.158 &10.667 &3.006  &-&-&-&- &-  &-&- &-    &-  &-  &-  &-      \\ 
CoGeTe$_3$       &  &6.751 &11.694 & 3.127  &-&-&-&- &-  &-&- &-    &-  &-  &-  &-   \\  \hline
NiGeS$_3$        &  &5.837 &10.106 &2.943  &-&-&-&-  &-  &-&- &-    &-  &-  &-  &-     \\ 
NiGeSe$_3$        &  &6.088 &10.543 &3.149   &6.089 &10.544  &3.149 &6.090 &10.548 & 3.169   &6.089 &10.5441  &3.149 & - &-  &-     \\ 
NiGeTe$_3$       & &6.450  &11.155 &3.334   &6.454 &11.162  &3.337 & - & - & - & - &- & -   &-  &-  &-      \\  \hline \hline
VSnS$_3$        &  &6.310 &10.209 &3.262   &6.288 &10.442  &3.164 &6.289 & 10.458 &3.199    &6.306 &10.473  &3.182  &6.274 &10.470  &3.133     \\ 
VSnSe$_3$       &  &6.625 &10.696 & 3.552 &6.655  &10.929  &3.462 &6.625 &10.955 & 3.458  &6.700  &11.000  &3.412  &6.623 &10.964  &3.431     \\ 
VSnTe$_3$        &  &7.146 &11.714 &3.775   &6.927 &11.876  &3.750 &6.984 & 12.015 & 3.686   & 6.879& 12.117 &3.620  &6.939 &11.875  &  3.722   \\  \hline
CrSnS$_3$       &  &6.240 &9.984 &3.224   &6.135 &10.627  &2.759 &6.107 & 10.578 &2.787  &6.152   &10.571  & 2.827  &6.095 & 10.648 &2.799     \\ 
CrSnSe$_3$       &  &6.504 &10.495 &3.529   &6.433 &11.144  &2.984 & 6.407 &11.102 &3.012    &6.477 &11.124  & 2.998  &6.409 &11.197  &  3.017    \\ 
CrSnTe$_3$       & &6.895  &11.642 &3.795 &6.914   &12.018  &3.409 &6.966 &12.112 & 3.406  &7.110  &12.239  &3.305 &7.106  & 12.378 &3.285     \\  \hline
MnSnS$_3$       &  &6.095 &10.557  & 2.535  &6.180 & 10.703  &2.830 &6.147 &10.648  & 2.643   &6.204 &10.662   &2.832  &6.148 & 10.609  &2.804    \\ 
MnSnSe$_3$       &  &6.437 &11.150 &2.718   &6.490 & 11.242 &3.055&6.491 &11.242 &3.003    &6.489 & 11.242 &3.064  &6.499 & 11.241 &3.075    \\ 
MnSnTe$_3$       &  &6.846 &11.450 &3.638   &7.019 &12.189  &3.314 &7.026 &12.136 & 3.341   &7.002 & 12.247 &3.313  &7.044 & 12.128  & 3.336    \\  \hline
FeSnS$_3$        &  &6.074 &10.519 &2.526   &6.164 &10.672  & 2.569 &10.715 &6.188 &  2.564  &6.163 &10.687  & 2.573 &6.223 & 10.822  & 2.571    \\ 
FeSnSe$_3$      &  &6.650 &11.515 &2.661   &6.548 &11.316  &2.832 &6.562 &11.364 &2.812    &6.529 &11.370  & 2.832 &6.577 & 11.312 &2.857     \\ 
FeSnTe$_3$      &  &6.764 &11.758 &3.148   &7.076 &12.259  & 3.158 &7.094 &12.293 &3.097    &7.051 &12.219  &3.115  &7.142 & 12.318  &3.095     \\  \hline
CoSnS$_3$       &  &6.001 &10.395 &2.560   &-&-&-&- &-  &-&-&-    &-  &-  &-  &-     \\ 
CoSnSe$_3$       & &6.331 &10.965 & 2.788   &-&-&-&- &-  &-&-&-    &-  &-  &-  &-     \\ 
CoSnTe$_3$        &   & 6.893 &11.940 &3.035   &-&-&-&- &-  &-&-&-    &-  &-  &-  &-    \\  \hline
NiSnS$_3$       &   &5.900 &10.218 &2.701  &-&-&-&- &-  &- &-&- &-  &-  &-  &-      \\ 
NiSnSe$_3$       &  &6.069 &10.507 & 3.013  &-&-&-&- &-  &-&-&-    &-  &-  &-  &-      \\ 
NiSnTe$_3$        &  & 6.309 &10.939 & 3.483 &-&-&-&- &-  &-&-&-    &-  &-  &-  &-    \\  \hline \hline
\end{tabular}  \label{tab:latticecons}
\end{center}
\end{table*}

\begin{table*}[htb!]
\caption{The theoretical two dimensional lattice parameters for 
transition metal trichalcogenide MAX$_3$ compounds.
These results were obtained using DFT-D2+U (=4eV) and are expressed in $\AA$; 
{\it a} and {\it b} are the in-plane lattice constants and c$'$ is the layer thickness,  
{\it i..e} it is the distance between the planes
containing the three chalcogen atoms in a single MAX$_3$ layer. In the cases of CoASe$_3$ and CoATe$_3$ (A=Ge, Sn), the U parameter value is 6 and 5 eV, respectively. In the case of CoSnS$_3$ the U parameter is 7eV.
}
\begin{center}
\begin{tabular}{cc|ccc|ccc|ccc|ccc|ccc}\\ \hline \hline
 MAX$_3$   &  & &{NM}&&&{ FM }&&& {nAFM}&&&{ zAFM }&&& {sAFM}& \\ \hline
                   &  &a(\AA) &b(\AA)& c$'$(\AA) &a(\AA)&b(\AA) & c$'$(\AA) &a(\AA) &b(\AA)   &c$'$(\AA)  &a(\AA)&b(\AA)  & c$'$(\AA) &a(\AA)&b(\AA) & c$'$(\AA) \\ \hline
VSiS$_3$   &  &6.022 &9.893 &3.399&6.041&10.461&3.179&6.033  &10.448  &3.184&6.042 &10.445    &3.193  & 6.032 &10.457  & 3.186 \\ 
VSiSe$_3$    &  &6.342 &10.454 &3.679&6.494&10.857&3.606&6.498  &11.250  &3.360&6.545 & 11.210   & 3.457 &6.514  &10.857  &3.533   \\ 
VSiTe$_3$    &  &6.888 &11.341 &4.030&6.777&11.865&3.922&7.183  &10.925  &4.090&6.758 &  11.592  & 3.938 &6.831  &11.578  & 3.940    \\  \hline
CrSiS$_3$    &  &5.865 &10.158 &3.044&5.959&10.322&3.194&5.930  &10.272  &3.216&5.968 &  10.265  & 3.238 &5.923  & 10.317 &3.220    \\ 
CrSiSe$_3$   &  &6.204 &10.745 &3.249&6.307&10.924&3.387&6.276  &10.871  &3.416&6.316 & 10.863   &3.435  &6.268  & 10.921 &3.416     \\ 
CrSiTe$_3$   &  &6.741 &11.672 &3.433&6.955&12.037&3.561&6.906  &11.959  &3.617&6.968 &  11.989  &3.619  & 6.970 & 11.984 &3.621    \\  \hline
MnSiS$_3$    &  &5.891 &10.207 &2.907&6.062&10.512&3.286&6.198  &10.736  &3.262&6.206 &  10.735  &3.294  &6.062  & 10.519 &3.301     \\ 
MnSiSe$_3$   &  &6.275 & 9.940 &3.568&6.364&11.033&3.491&6.342  & 10.990 &3.516& 6.358&  11.018  &3.596  &6.331  & 11.044 &3.521     \\ 
MnSiTe$_3$    &  &6.803 &11.783 &3.227&6.905&11.955&3.721&7.020  &11.667  &3.775&6.801 &  11.943  &3.887  &6.878  & 12.090 & 3.732    \\  \hline
FeSiS$_3$    &  &6.039 &10.474 &2.831&6.005&10.405&3.269&  6.014 &10.418  &3.257&6.018 &   10.448 & 3.254 &6.009  & 10.415 & 3.258    \\ 
FeSiSe$_3$    &  &6.376 &11.067 &3.011&6.353&11.000&3.478& 6.361 &11.013  &3.442& 6.335&  10.987  & 3.443 &6.351  & 11.016 & 3.451    \\ 
FeSiTe$_3$    &  &6.521 &11.283 &3.482&6.932&12.001&3.753&6.932  &12.001 &3.678 &6.890 & 12.047   &3.641 &6.932  &12.001& 3.748  \\  \hline
CoSiS$_3$    &  &5.737 &9.936 &2.901&5.900&10.226&3.071&5.931  & 10.272 &3.032&5.963 &  10.289  &3.068  &5.929  & 10.230 &3.069     \\ 
CoSiSe$_3$    &  &6.071 &10.515 &3.064&6.150&10.649&3.195&6.148  & 10.650 &3.193& 6.151& 10.652   &3.204  &6.141  &10.654  & 3.185   \\ 
CoSiTe$_3$    &  &6.608 &11.444 &3.187&6.723&11.617&3.282&6.699  &  11.599&3.283& 6.692&  11.584  &3.306  &6.715  &11.650  & 3.280   \\  \hline
NiSiS$_3$    &  &5.719 &9.888 &3.001&5.719&9.912&3.079&5.741  &9.947  &3.054&5.741 & 9.922    &3.186  &5.740  &9.936  &3.065     \\ 
NiSiSe$_3$   &  &6.030 &10.430 &3.163&6.032&10.441&3.241&6.033  &10.449  &3.226&6.033 &10.458    &3.232  &6.033  &10.458  &3.232     \\ 
NiSiTe$_3$    &  &6.575 &11.374 &3.290&6.558&11.358&3.381&6.564  &11.366  &3.346&6.590 &11.348    &3.365  &6.574  & 11.365 &3.356     \\  \hline \hline
VGeS$_3$    &  &6.161 &10.168 &3.302&6.133&10.624&3.183&6.124  & 10.608 &3.191&6.132 &10.603    &3.198  & 6.124 &10.617  &3.194    \\ 
VGeSe$_3$    &  &6.424 &10.558 &3.696&6.459&11.181&3.379&6.561  & 11.363 &3.380&6.458 &11.165    & 3.396  &6.530  &11.443  & 3.339    \\ 
VGeTe$_3$   &  &6.940 &11.614 &3.888&7.225&10.899&4.055&7.225  &10.899  &4.166&7.235 &10.900    &4.180  &7.538  &10.791  & 4.108    \\  \hline
CrGeS$_3$    &  &5.969 &10.386 &2.997&6.047&10.474&3.199&6.016  &10.421  & 3.224&6.051 & 10.415   &3.243  & 6.011 &10.464  & 3.230    \\ 
CrGeSe$_3$   &  &6.311 &10.932 &3.277&6.384&11.058&3.397&6.350  &11.001  &3.428&6.387 &10.992    &3.447  & 6.344 &11.046  &3.431     \\ 
CrGeTe$_3$    &  &7.030 &11.156 &3.817&7.029&12.178&3.564&6.983  &12.100  &3.621&7.026 &12.051    &3.637  &6.978  &12.162  &3.614      \\  \hline
MnGeS$_3$    &  &6.032 &10.448 &2.593&6.130&10.627&3.314& 6.131 & 10.618 &3.320&6.278 &  10.863  &3.303  &6.269  &10.875  & 3.288      \\ 
MnGeSe$_3$    &  &6.328 &10.962 &3.056&6.430&11.140&3.511&6.402  &11.097  & 3.541&6.435 & 11.098   &3.586  &6.399  & 11.166 & 3.537    \\ 
MnGeTe$_3$   &  &6.909 &11.958 &3.190&7.038&12.174& 3.190&7.003  &  12.154& 3.738& 6.862&  12.062  & 3.883  &6.922  &  11.925&  3.802    \\  \hline
FeGeS$_3$     &  &6.075 &10.522 &2.789&6.117&10.553&3.281&6.103  & 10.576 &3.263& 6.103&  10.576  &3.271  & 6.103 & 10.587 &3.267     \\ 
FeGeSe$_3$    &  &6.404 &11.096 &2.973&6.434&11.119& 3.467&6.425  & 11.133 &3.485& 6.438&11.254    &3.436  &6.421  &11.134  & 3.460    \\ 
FeGeTe$_3$     &  &6.586 &11.485 &3.526&6.933&12.069&3.766&6.980  &12.076  &3.654&6.917 & 12.106   &3.689  &6.997  &12.127  &3.646    \\  \hline
CoGeS$_3$    &  &5.825 &10.090 &2.878&5.971&10.350&3.064&5.996  & 10.385 &3.029&6.009 & 10.397   &3.056  &5.999  &10.378  &3.067     \\ 
CoGeSe$_3$      &  &6.147 &10.647 & 3.049&6.146&10.646&3.049&6.217  &10.768  &3.171&6.218 & 10.771   &3.200  &6.146  &10.646  & 3.049     \\ 
CoGeTe$_3$      &  &6.666 & 11.550&3.176&6.749&11.664& 3.277&6.730  & 11.664 & 3.256&6.774 &  11.728  &3.291  &6.739  &11.686  &3.158     \\  \hline
NiGeS$_3$        &  &5.807 & 10.050&2.963&5.837&10.113&3.047&5.845  & 10.128 &3.033&5.852 &  10.128  &  3.116 & 5.852 & 10.128 &  3.069     \\ 
NiGeSe$_3$        &  &6.105 & 10.571&3.129&6.094&10.550&3.242&6.098  & 10.562 &3.222&6.107 &  10.570  &3.221  &6.107  &10.570  & 3.221     \\ 
NiGeTe$_3$       &  &6.457 &11.169 &3.313&6.658&11.523&3.176&6.656  &  11.514& 3.185&- &-    & - &-  &  -&  -    \\  \hline \hline
VSnS$_3$        &  &6.236 &10.805 &2.736&6.308&10.926&2.813&6.303  &  10.918&2.815&6.328 & 10.887   & 2.869 &6.293  &10.948  & 2.824    \\ 
VSnSe$_3$       &  &6.645 &10.597 &3.699&6.384&11.058&3.550  &6.713&11.285  & 3.349& 6.734&11.266    &3.362  &6.711  &11.308  &3.328      \\ 
VSnTe$_3$        &  &7.255 & 11.608&3.815&7.151&11.628&4.203&7.512  &11.090  &4.279& 6.833& 12.110   &4.184  &6.833  &12.110  &4.189     \\  \hline
CrSnS$_3$       &  &6.276 &10.005 & 3.239&6.216&10.766&2.796& 6.191 & 10.724  &2.818& 6.223& 10.727   & 2.824  &6.188  &10.769  & 2.820     \\ 
CrSnSe$_3$     &  &6.384 &11.058& 3.141&6.537&11.314&3.026&6.504  &11.258  &3.060&6.545 &11.280    &3.044  &6.506  &11.323  & 3.047     \\ 
CrSnTe$_3$      &  &7.055 &11.847 &3.593&7.150&12.391& 3.661&7.126&12.344&3.711&7.127  &12.269  &3.743&7.102 &12.343    & 3.726    \\  \hline
MnSnS$_3$     &  &6.160 &10.669 &2.595&6.250&10.822&2.903&6.690  & 10.422 & 3.094&6.699 &10.425    &3.098  &6.190  &10.932  &   2.952   \\ 
MnSnSe$_3$      &  &6.501 &11.257 &2.800&6.517&11.287&3.141&6.471  &11.721  &3.476&6.503 & 11.274   &3.228  & 6.452 &11.757  &3.469     \\ 
MnSnTe$_3$       &  &6.810 &11.463 &3.611&7.074&12.264& 3.849&7.063  &12.219  &3.891& 7.063&  12.231  & 3.938  & 7.136 &  12.343&  3.869    \\  \hline
FeSnS$_3$       &  &6.182 &10.703 &2.515&6.246&10.817&2.888&6.219  & 10.770 &2.875&6.223 &  10.817  &2.888  & 6.235 & 10.772 &  2.904    \\ 
FeSnSe$_3$      &  &6.616 &11.261 &3.150&6.483&11.237&3.115&6.483  & 11.237 &3.115&6.488 &  11.298  & 3.195  &6.513  & 11.237 &  3.204    \\ 
FeSnTe$_3$      &  &6.646 & 11.681&3.244&7.241&12.434&3.562& 7.241 & 12.434 &3.562& 7.241&   12.434 & 3.556 & 7.241 &  12.434& 3.562     \\  \hline
CoSnS$_3$       &  &5.986 &10.369 &2.594&6.044&10.470&2.644& 5.987 &  10.370&2.594& 6.045&   10.452 &2.647  & 6.045 & 10.452 &2.651      \\ 
CoSnSe$_3$       &  &6.297 &10.905 &2.802&6.357&11.008&2.858& 6.291 & 10.897 &2.803& 6.369&  10.997  &2.883  &6.341  &11.004  & 2.883    \\ 
CoSnTe$_3$       &  &6.840 & 11.861&3.179&6.864&11.859&3.120&  6.920&  11.987& 3.023& 6.893&   11.904 &3.201  &6.888  & 11.945 &   3.165    \\  \hline
NiSnS$_3$       &  &5.902 & 10.224&2.695&5.904&10.237&2.745&  5.875&  10.462&2.849& 5.910   & 10.245 &2.747  &5.894  & 10.247 &2.782     \\ 
NiSnSe$_3$      &  &6.077 & 10.519&2.998&6.147&10.646& 2.998&6.083  &  10.529&2.994&6.110    &10.633  &3.015  &6.153  &10.696 & 2.960     \\ 
NiSnTe$_3$        &  &6.317 & 10.954&3.462&6.404&11.145& 3.462&-  &-  &-&- &-    &-  &-  &-  &-    \\  \hline \hline
\end{tabular}  \label{tab:bulklatticecons}

\end{center}
\end{table*}

\begin{figure*}[htbp!]
\begin{center}
\includegraphics[width=15cm]{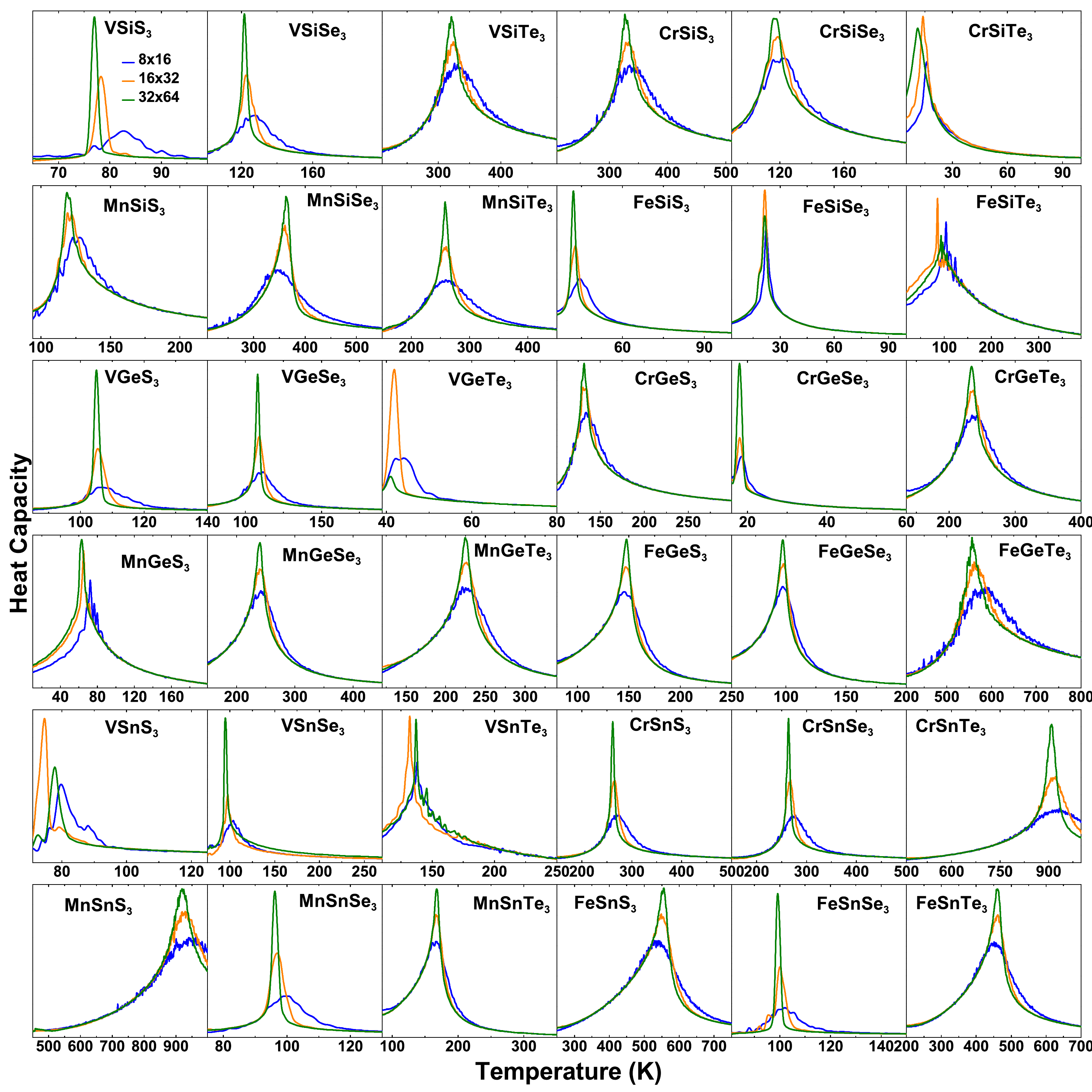}
\caption{(Color online) 
Temperature dependent evaolution of the heat capacity calculated with the Metropolis Monte Carlo algorithm in a Lx2L (L=8, 16, 32) superlattice for the effective Ising model, calculated using three nearest neighbour J-parameters obtained from the DFT-D2 total energies. The transition temperature is estimated from the maximum in the curve.
}
\label{Tc}
\end{center}
\end{figure*}

\begin{figure*}[htbp!]
\begin{center}
\includegraphics[width=15cm]{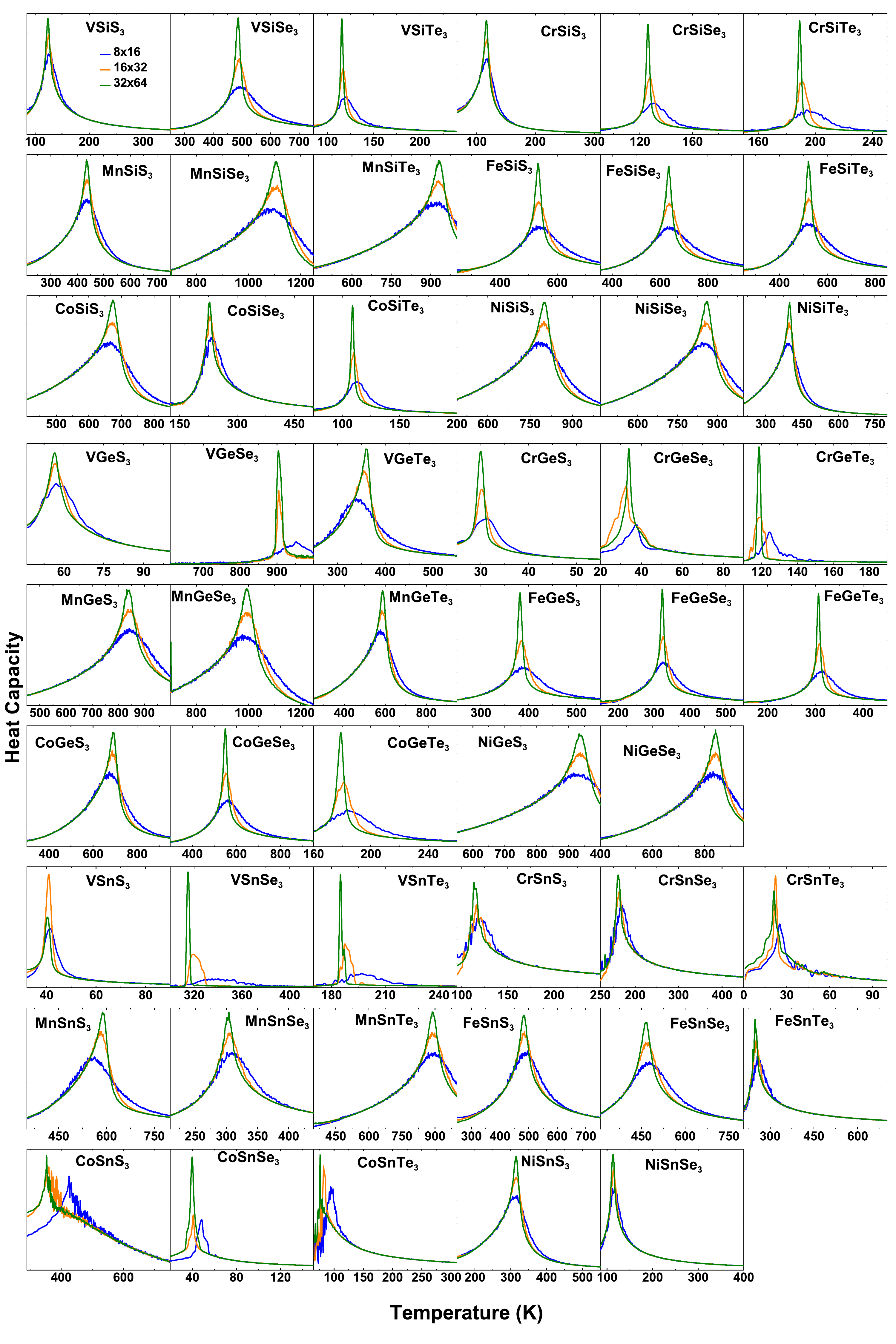}
\caption{(Color online) 
Temperature dependent evaolution of the heat capacity calculated with the Metropolis Monte Carlo algorithm in a  Lx2L (L=8, 16, 32) superlattice for the effective Ising model, calculated using three nearest neighbour J-parameters obtained from the DFT-D2+U (=4eV) total energies.  In the cases of CoASe$_3$ and CoATe$_3$ (A=Ge, Sn), the U parameter value is 6 and 5 eV, respectively. In the case of CoSnS$_3$ the U parameter is 7eV. The transition temperature is estimated from the maximum in the curve.
}
\label{Tc_U}
\end{center}
\end{figure*}

\begin{table*}[htb!]
\caption{The critical temperatures (T$_C$) of MAX$_3$ compounds obtained from the heat capacity calculated through the Metropolis Monte Carlo algorithm in a Lx2L (L=8, 16, 32) superlattice for the effective Ising model using three nearest neighbour J-parameters obtained from the total energies. The critical temperatures are listed in Kelvin (K) units and have values that 
depend substantially on the exchange-correlation approximation employed within DFT-D2 and DFT-D2+U (=4eV). In the cases of CoASe$_3$ and CoATe$_3$ (A=Ge, Sn), the U parameter value is 6 and 5 eV, respectively. 
In the case of CoSnS$_3$ the U parameter is 7eV. Different methods have been indicated by I:DFT-D2 and II:DFT-D2+U. 
}
\begin{center}
\begin{tabular}{cc|ccc|ccc|ccc}\\ \hline \hline
A&&   \multicolumn{3}{c|}{Si}  &  \multicolumn{3}{c|}{Ge}  &  \multicolumn{3}{c}{Sn}  \\ \hline
MAX$_3$&Method&{L=8}&{L=16}&{L=32}&{L=8}&{L=16}&{L=32}&{L=8}&{L=16}&{L=32}\\ \hline
VAS$_3$&I           &83 &78 &77 &107 &105 &105 &80 &75 &78 \\
              &II          &125 &125 &123 &57 &57 &56 &41 &40 &40 \\ \hline
VASe$_3$&I           &127 &123 &122 &111 &109 &108 &104 &98 &95 \\
              &II          &494 &490 &487 &952 &905 &903 &338 &319 &314 \\ \hline
VATe$_3$&I           &326 &324 &321 &42 &42 &41 &138 &132 &137 \\
              &II          &119 &116 &115 &339 &354 &360 &196 &187 &185 \\  \hline \hline
CrAS$_3$&I           &339 &333 &326 &134 &130 &132 &269 &266 &262 \\
              &II          &116 &116 &115 &31 &30 &30 &119 &115 &113 \\ \hline
CrASe$_3$&I           &119 &119 &116 &19 &18 &18 &278 &268 &264 \\
              &II          &130 &127 &126 &37 &32 &34 &170 &162 &160 \\ \hline
CrATe$_3$&I           &16 &14 &11 &235 &234 &234 &933 &921 &909 \\
              &II          &194 &190 &189 &124 &118 &118 &25 &22 &21 \\  \hline \hline
MnAS$_3$&I           &123 &119 &119 &72 &65 &65 &962 &929 &919 \\
              &II          &436 &436 &436 &843 &843 &837 &562 &580 &588 \\ \hline
MnASe$_3$&I           &352 &358 &362 &241 &239 &239 &99 &97 &96 \\
              &II          &1085 &1103 &1103 &977 &993 &993 &309 &305 &304 \\ \hline
MnATe$_3$&I           &259 & 257&258 &225 &226 &224 &166 &166 &168 \\
              &II          &928 &928 &930 &579 &586 &586 &895 &891 &891 \\  \hline \hline
FeAS$_3$&I           &44 &43 &42 &145 &147 &148 &540 &549 &556 \\
              &II          &529 &534 &527 &388 &384 &382 &486 &484 &483 \\ \hline
FeASe$_3$&I           &22 &21 &21 &97 &98 &97 &102 &100&99 \\
              &II          &635 &638 &631 &325 &325 &322 &475&472 &466 \\ \hline
FeATe$_3$&I           &104 &86 &86 &576 &561 &556 &451 &462 &461 \\
              &II          &516 &520 &518 &314 &309 &307 &256 &252 &249 \\  \hline \hline
CoAS$_3$&I           &- &- &- &- &- &- &- &- &- \\
              &II          &665 &672 &672 &674 &686 &691 &427 &359 &351 \\ \hline
CoASe$_3$&I           &- &- &- &- &- &- &- &- &- \\
              &II          &232 &229 &227 &563 &554 &549 &48 &41 &39 \\ \hline
CoATe$_3$&I           &- &- &- &- &- &- &- &- &- \\
              &II          &112 &110 &109 &184 &181 &178 &93 &82 &76 \\  \hline \hline
NiAS$_3$&I           &- &- &- &- &- &- &- &- &- \\
              &II          &112 &110 &109 &917 &935 &935 &312 &316 &316 \\ \hline
NiASe$_3$&I           &- &- &- &- &- &- &- &- &- \\
              &II          &844 &851 &861 &833 &843 &843 &116 &115 &113 \\ \hline
NiATe$_3$&I           &- &- &- &- &- &- &- &- &- \\
              &II          &392 &398 &398 &- &- &- &- &- &- \\
\hline \hline
\end{tabular} \label{tab:gaps}
\end{center}
\end{table*}

\begin{figure*}[htb!]
\begin{center}
\includegraphics[width=11cm]{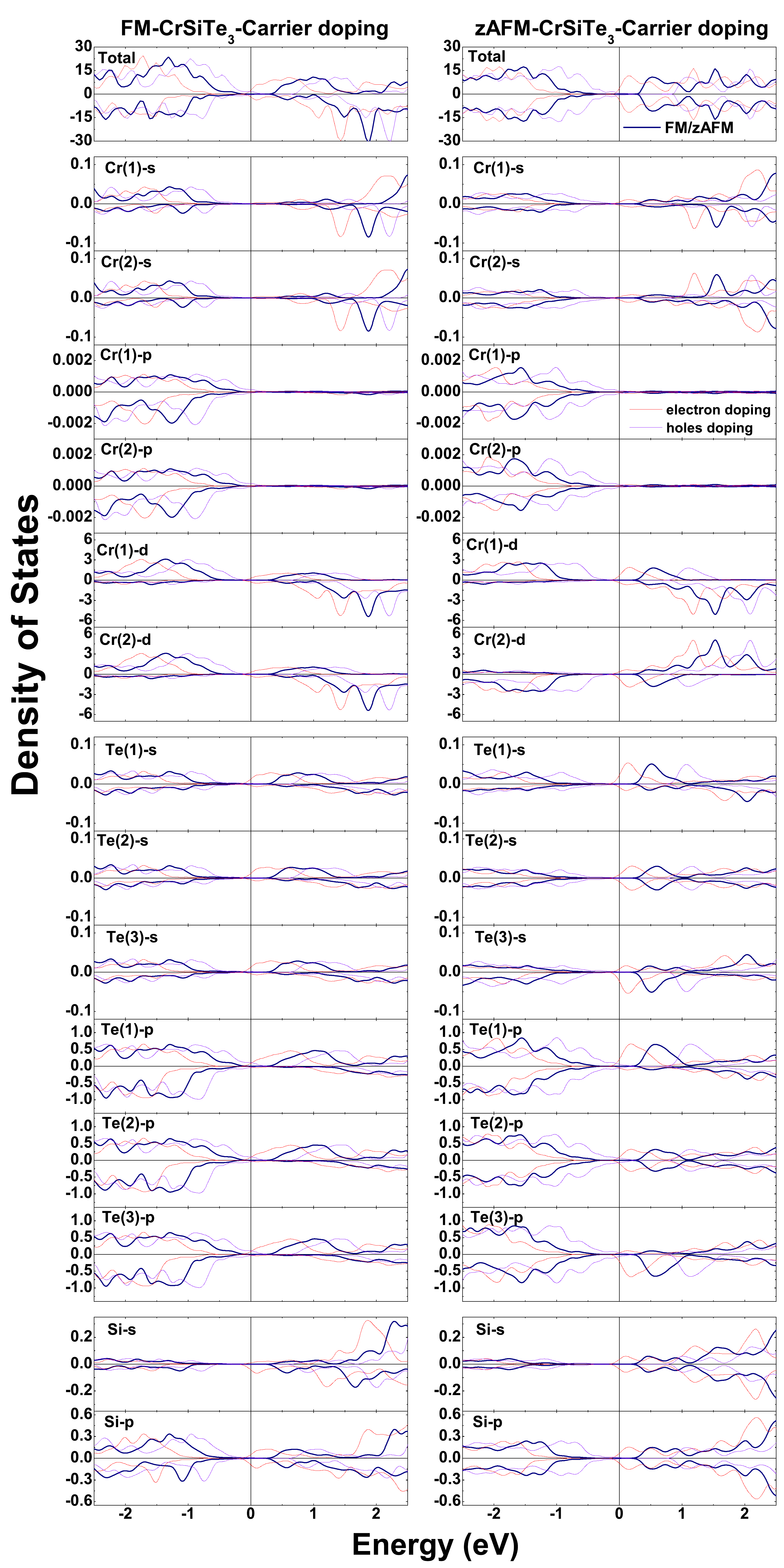}
\caption{(Color online) Comparison of the density of states (DOS) and projected DOS of carrier doping of -0.2 (holes per formula unit) and 0.2 (electrons per formula unit) in CrSiTe$_3$ between FM and zAFM states. } 
\label{fig:carrier-dos}
\end{center}
\end{figure*}

\begin{figure*}[htb!]
\begin{center}
\includegraphics[width=11cm]{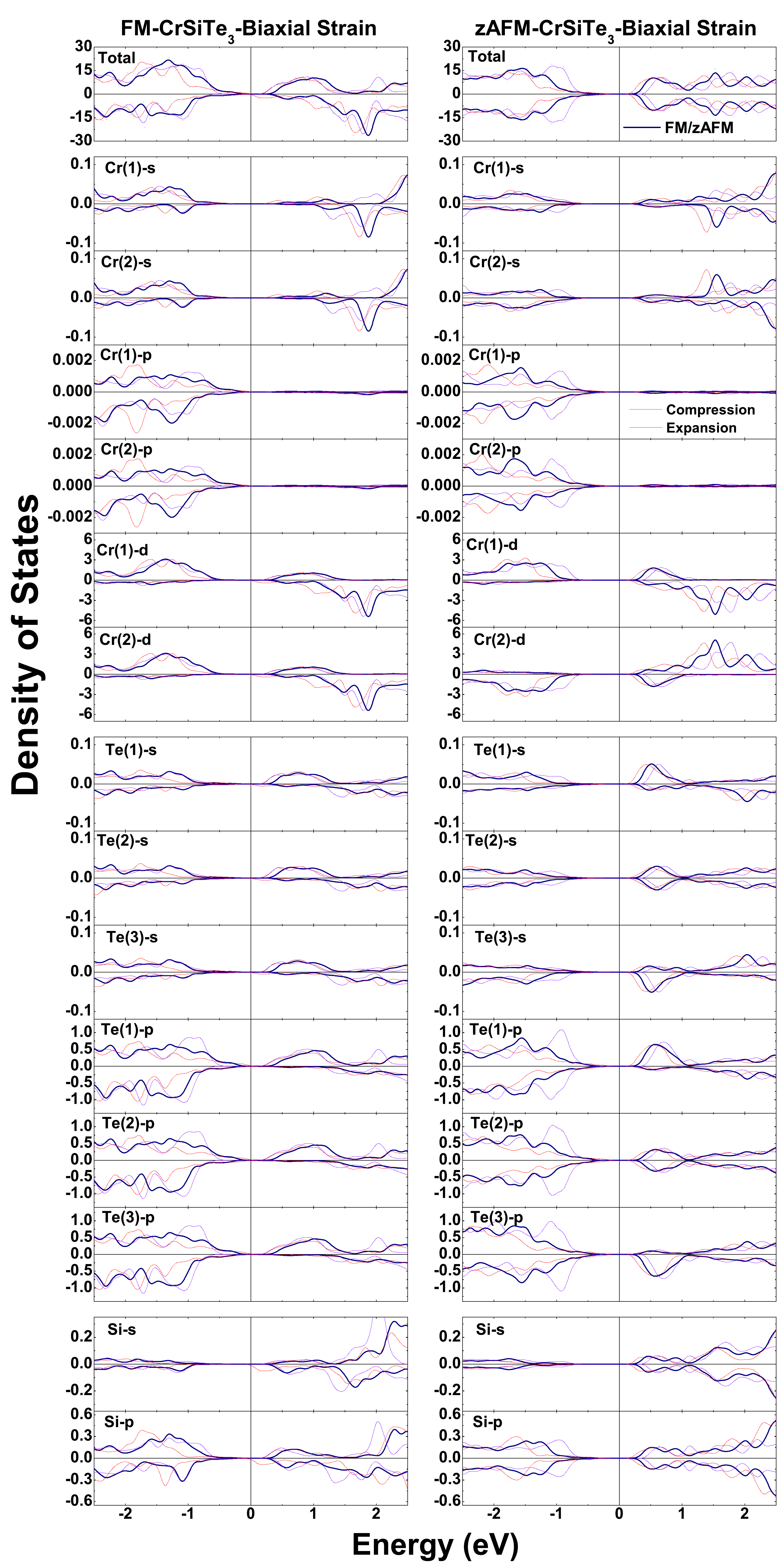}
\caption{(Color online) Comparison of the density of states (DOS) and projected DOS of biaxial strains of -4{\%} (compressive) and 4{\%} (expansive) in CrSiTe$_3$ between FM and zAFM states. } \label{fig:str-dos}
\end{center}
\end{figure*}
%
%

\begin{table}[htb!]
\caption{ The comparision of magnetic moments (MM) in Bohr magneton $\mu _B$ per metal atom for single layer CrSiTe$_3$, 
the three nearest neighbour exchnage coupling strengths {\it J$_i$} in meV, and Monte Carlo estimates (for 32x64) of the 
critical temperatures in Kelvin for applied  biaxial strains of -4{\%} (compressive) and 4{\%} (expansive), 
as well as carrier doping of -0.2 (holes per formula unit) and 0.2 (electrons per formula unit)}
\begin{center}
\begin{tabular}{cccccc}\\ \hline \hline 
CrSiTe$_3$                                & MM    &{\it J$_1$} &{\it J$_2$} &{\it J$_3$} & T$_C$   \\ \hline
Neutral                                      &2.851 & -0.894     &0.250          & 0.026        & 16 \\
4{\%} (expansive)                   &3.188  & -1.114     &1.312          & -0.371        & 104 \\
-4{\%} (compressive)               &3.022  & 0.404     &-0.781          & 0.135        & 476 \\
0.2 (electrons per formula unit)&3.100  & -2.875     &1.607          & 2.161        & 720 \\
 -0.2 (holes per formula unit)    &3.092 & 1.622     &-1.116          &-2.603        & 668 \\ \hline \hline
\end{tabular}  \label{tab:mae}
\end{center}
\end{table}

\end{document}